\documentclass[prd,twocolumn,superscriptaddress,showpacs,letterpaper,nofootinbib,aps]{revtex4-1}

\usepackage{color}
\usepackage{graphicx}
\usepackage{amsmath}
\usepackage{times}
\usepackage{bm}
\usepackage{acronym}
\usepackage{subfigure}
\usepackage[bookmarksnumbered, bookmarksopen, breaklinks, colorlinks]{hyperref}

\def\be{\begin{equation}}
\def\ee{\end{equation}}

\def\bea{\begin{eqnarray}}
\def\eea{\end{eqnarray}}

\def\Msun{M_\odot}

\begin{document}

\title{Accuracy of gravitational waveform models for observing neutron-star--black-hole binaries in Advanced LIGO}

\author{Alexander H.\ Nitz}
\affiliation{Department of Physics, Syracuse University, Syracuse NY 13244}

\author{Andrew Lundgren}
\affiliation{Albert-Einstein-Institut, Max-Planck-Institut f\"ur
Gravitationsphysik, D-30167 Hannover, Germany}
\affiliation{Leibniz Universit\"at Hannover, D-30167 Hannover, Germany}
\affiliation{Kavli Institute of Theoretical Physics, University of California, Santa Barbara, CA 93106}

\author{Duncan A.\ Brown}
\affiliation{Department of Physics, Syracuse University, Syracuse NY 13244}
\affiliation{Kavli Institute of Theoretical Physics, University of California, Santa Barbara, CA 93106}
\affiliation{LIGO Laboratory, California Institute of Technology, Pasadena CA 91125}

\author{Evan Ochsner}
\affiliation{Center for Gravitation and Cosmology, University of Wisconsin-Milwaukee, Milwaukee, WI 53211}
\affiliation{Kavli Institute of Theoretical Physics, University of California, Santa Barbara, CA 93106}

\author{Drew Keppel}
\affiliation{Albert-Einstein-Institut, Max-Planck-Institut f\"ur
Gravitationsphysik, D-30167 Hannover, Germany}
\affiliation{Leibniz Universit\"at Hannover, D-30167 Hannover, Germany}

\author{Ian W.\ Harry}
\affiliation{Department of Physics, Syracuse University, Syracuse NY 13244}
\affiliation{Kavli Institute of Theoretical Physics, University of California, Santa Barbara, CA 93106}


\begin{abstract}
Gravitational waves radiated by the coalescence of compact-object binaries
containing a neutron star and a black hole are one of the most interesting
sources for the ground-based gravitational-wave observatories Advanced LIGO and
Advanced Virgo. Advanced LIGO will be sensitive to the inspiral of a $1.4\, M_\odot$
neutron star into a $10\,M_\odot$ black hole to a maximum distance of $\sim 900$~Mpc. 
Achieving this sensitivity and extracting the physics imprinted in
observed signals requires accurate modeling of the binary to construct
template waveforms. In a neutron star--black hole binary, the black hole may
have significant angular momentum (spin), which affects the phase evolution of
the emitted gravitational waves. We investigate the ability of currently available
post-Newtonian templates to model the gravitational waves emitted during the
inspiral phase of neutron star--black hole binaries. We restrict to the case where the spin of the
black hole is aligned with the orbital angular momentum and compare several
post-Newtonian approximants. We examine
restricted amplitude post-Newtonian waveforms that are accurate to
third-and-a-half post-Newtonian order in the orbital dynamics and complete to second-and-a-half post-Newtonian order
in the spin dynamics. We also consider post-Newtonian waveforms that include the recently derived third-and-a-half
post-Newtonian order spin-orbit correction and the third post-Newtonian order spin-orbit tail correction. 
We compare these post-Newtonian approximants to the effective-one-body waveforms for spin-aligned binaries.
For all of these waveform families, we find that
 there is a large disagreement between
different waveform approximants starting at low to moderate black hole spins,
particularly for binaries where the spin is anti-aligned with the orbital
angular momentum. The match between the TaylorT4 and TaylorF2 approximants is $\sim 0.8$ for a binary with $m_{BH}/m_{NS} \sim 4$ and 
$\chi_{BH} = cJ_{BH}/Gm^2_{BH} \sim 0.4$.
We show that the divergence between the gravitational waveforms begins in the early
inspiral at $v \sim 0.2$ for $\chi_{BH} \sim 0.4$.  Post-Newtonian spin corrections beyond those currently
known will be required for optimal detection searches and to measure the
parameters of neutron star--black hole binaries. The strong dependence of 
the gravitational-wave signal on the spin dynamics will make it possible to extract significant
astrophysical information from detected systems with Advanced LIGO and
Advanced Virgo.
\end{abstract}

\maketitle

\acrodef{aLIGO}[aLIGO]{Advanced Laser Interferometer Gravitational-wave Observatory}
\acrodef{AdV}[AdV]{Advanced Virgo}
\acrodef{LIGO}[LIGO]{Laser Interferometer Gravitational-wave Observatory}
\acrodef{CBC}[CBC]{compact binary coalescence}
\acrodef{S6}[S6]{LIGO's sixth science run}
\acrodef{VSR23}[VSR2 and VSR3]{Virgo's second and third science runs}
\acrodef{EM}[EM]{electromagnetic}
\acrodef{NS}[NS]{neutron star}
\acrodef{BH}[BH]{black hole}
\acrodef{BNS}[BNS]{binary neutron star}
\acrodef{NSWD}[NSWD]{neutron star-white dwarf}
\acrodef{NSBH}[NSBH]{neutron star and a black hole}
\acrodef{GRB}[GRB]{gamma-ray burst}
\acrodef{S5}[S5]{LIGO's fifth science run}
\acrodef{S4}[S4]{LIGO's fourth science run}
\acrodef{VSR1}[VSR1]{Virgo's first science run}

\acrodef{PSD}[PSD]{power spectral density}
\acrodef{VSR3}[VSR3]{Virgo's third science run}
\acrodef{BBH}[BBH]{binary black holes}
\acrodef{SNR}[SNR]{signal-to-noise ratio}
\acrodef{SPA}[SPA]{stationary-phase approximation}
\acrodef{LHO}[LHO]{LIGO Hanford Observatory}
\acrodef{LLO}[LLO]{LIGO Livingston Observatory}
\acrodef{LSC}[LSC]{LIGO Scientific Collaboration}
\acrodef{PN}[PN]{post-Newtonian}
\acrodef{DQ}[DQ]{data quality}
\acrodef{IFO}[IFO]{interferometer}
\acrodef{DTF}[DTF]{detection template families}
\acrodef{FAR}[FAR]{false alarm rate}
\acrodef{FAP}[FAP]{false alarm probability}
\acrodef{PTF}[PTF]{physical template family}
\acrodef{ADE}[ADE]{advanced detector era}
\acrodef{FFT}[FFT]{Fast Fourier Transformation}
\acrodef{GPU}[GPU]{graphical processing unit}
\acrodef{ISCO}[ISCO]{inner-most stable circular orbit}
\acrodef{MECO}[MECO]{minimum energy circular orbit}

\section{Introduction}
\label{sec:introduction}

Compact object binaries are likely to be the first source detected by the
\ac{aLIGO}~\cite{Harry:2010zz} and Advanced Virgo (AdV)~\cite{aVirgo}. These detectors will be sensitive to the
gravitational waves radiated as the orbital frequency of the binary sweeps
upwards from $\sim 5$--$10$~Hz to the point at which the compact objects
coalesce~\cite{Th300}. 
Binaries containing a \ac{NSBH} have a predicted 
coalescence rate of $0.2$--$300\ \textrm{yr}^{-1}$ within the sensitive volume
of aLIGO~\cite{Abadie:2010cf}, making them an important source for these
observatories. The observation of a \ac{NSBH} by \ac{aLIGO} would be the first 
conclusive detection of this class of compact-object binary.
Gravitational-wave observations of \ac{NSBH} binaries will allow us to explore the central engine of short,
hard gamma-ray bursts, shed light on models of stellar evolution and core
collapse, and investigate the dynamics of compact objects in the strong-field regime~\cite{lrr-2009-2, Eichler:1989ve, Narayan:1992iy, Paczynski:1991aq, Berger:2010qx, Fryer:2011cx, Hannam:2013uu}.
Achieving aLIGO's optimal sensitivity to
\ac{NSBH} binaries and exploring their physics 
requires accurate modeling of the gravitational waves emitted 
over many hundreds of orbits as the signal sweeps through the detector's
sensitive band. For \ac{BNS} systems the mass
ratio between the two neutron stars is small and the angular momenta of the
neutron stars (the neutron stars' spins) is low. In this case, the emitted waves are
well modeled by \ac{PN}
theory~\cite{Blanchet:2006zz,Buonanno:2009zt,Brown:2012qf}. 
However, \ac{NSBH} binaries can have significantly larger mass ratios and the spin of
the black hole can be much larger than that of a neutron star. The combined
effects of mass ratio and spin present challenges in constructing accurate gravitational waveform models for
\ac{NSBH} systems, compared to \ac{BNS} systems.  In this paper we
investigate how accurately current theoretical models simulate \ac{NSBH} gravitational waveforms
within the sensitive frequency band of \ac{aLIGO}.

Although no \ac{NSBH} binaries have been directly observed, both \acp{BH} and \acp{NS}
have been observed in other binary systems. Several \ac{BNS} systems and
\ac{NSWD} systems have been observed by detecting their electromagnetic
signatures. Electromagnetic observations suggest that the \ac{NS} mass
distribution in \ac{BNS} peaks at $1.35 \Msun$--$1.5 \Msun$ with a narrow
width~\cite{Kiziltan:2010ct}, although \acp{NS} in globular clusters seem to
have a considerably wider mass distribution~\cite{Kiziltan:2010ct}.  There is
also evidence that a neutron star in one system has a mass as high as $\sim 3
\Msun$~\cite{Freire:2007jd}.  The dimensionless spin magnitude $\chi = cJ/Gm^2$ for
\acp{NS} is constrained by possible \ac{NS} equations of state to a maximum of
0.7~\cite{Lo:2010bj}.  The fastest observed pulsar has a spin period
of 1.4 ms~\cite{Hessels:2006ze}, corresponding to a $\chi \sim 0.4$, and the
most rapidly spinning observed \ac{NS} in a binary, J0737--3039A, has a spin
of only $\chi \sim 0.05$.  The observational data for \acp{BH} is more limited
than for \acp{NS}.  Studies of \acp{BH} in low-mass X-ray
binaries suggest a mass distribution of $7.8 \pm 1.2 \Msun$~\cite{Ozel:2010su}. This extends to $8-11 \pm 2-4 \Msun$ when 5
high-mass, wind-fed, X-ray binary systems are included~\cite{Farr:2010tu}. For
\acp{BH} there is evidence for a broad distribution of spin
magnitudes~\cite{Miller:2009cw}, although general relativity limits it to be
$\chi < 1$. Given the uncertainties in the masses and spins of
\ac{NSBH} binaries, we consider a fairly broad mass and spin
distribution when investigating the accuracy of \ac{NSBH} waveforms.
In this paper, we consider \ac{NSBH} binaries with the \ac{NS} mass between 1
and $3\, M_\odot$, the \ac{BH} mass between $3$ and $15\, M_\odot$, the
\ac{NS} spin between 0 and $0.05$ and the
\ac{BH} spin between 0 and 1. Between these limits, the distributions of mass and spin are all
assumed to be uniform. 

Gravitational-wave detectors are sensitive to the phase evolution of the waves radiated
by the binary. \ac{PN} theory can be used to compute the
energy of a compact binary $E(v)$ and the flux radiated in gravitational waves
$\mathcal{F}(v)$ in terms of the invariant velocity $v = (\pi M f)^{1/3}$,
where $M = m_1 + m_2$ is the total mass of the binary, and $f$ is the
gravitational-wave frequency~\cite{Blanchet:2006zz}. By solving the energy
balance equation $dE/dt = - \mathcal{F}$, we can obtain expressions for the
gravitational-wave phase as a function of time $\phi(t)$ or, equivalently, the
Fourier phase of the waves as a function of frequency $\Psi(f)$. At leading
order, the gravitational wave phase depends only on the chirp mass
$\mathcal{M}_c = (m_1 m_2)^{3/5} / (m_1 + m_2)^{1/5}$~\cite{Peters:1963ux}.
Beyond leading order, the waveforms also depend on the symmetric mass
ratio $\eta = m_1 m_2 / (m_1 + m_2)^2$~\cite{Wiseman:1993aj,Blanchet:1995fg,Blanchet:1995ez,Blanchet:1996pi,
Blanchet:2001ax, Blanchet:2004ek}, with spin-orbit 
corrections entering at the third correction beyond leading order~\cite{Kidder:1992fr,
Kidder:1995zr,Arun:2008kb,Blanchet:2012sm,Bohe:2013cla}.

There are several different ways in which to solve the energy balance equation
to obtain the gravitational-wave phase measurable by aLIGO; these different methods are known as
\ac{PN} \emph{approximants.} While the convergence of the full \ac{PN} series 
is not guaranteed, the available \ac{PN} approximants produce waveforms that are indistinguishable 
for \ac{BNS} systems in Advanced LIGO and are reliable for use in detection searches and parameter
measurement~\cite{Simone:1996db,Buonanno:2009zt,Brown:2012qf}. However, for \ac{NSBH}
binaries the total mass, and hence the \ac{PN} expansion parameter $v$, is
larger. The mass ratio and spin corrections are also more significant.
In this paper, we investigate the accuracy of 
waveforms generated by different \ac{PN} approximants for observing \ac{NSBH}
binaries with aLIGO.
To do this, one could compare subsequent terms in the \ac{PN} expansion and
determine the effect of neglecting them. However, in the case of systems whose
component objects are spinning, only terms up to 2.5\ac{PN} order are
completely known~\cite{Kidder:1992fr,Kidder:1995zr,Arun:2008kb}. 
This represents the leading order (1.5\ac{PN}) and
next-to-leading order (2.5\ac{PN}) spin-orbit, along with the leading order
(2.0\ac{PN}) spin-spin contributions to the phasing~\cite{Kidder:1992fr,Kidder:1995zr,Arun:2008kb}.  
We choose to compare approximants that are constructed with terms up to the same
\ac{PN} order, but that use inversely related differential equations to solve
for the orbital dynamics, in addition to comparing to approximants that include
higher order spin-related corrections at partially derived orders~\cite{Bohe:2013cla, Blanchet:2011zv}.
These methods both have the effect of testing
how well the \ac{PN} series has converged. We also present a comparison between
waveforms from these \ac{PN} approximants where we fix the mass and spin
parameters of the objects in order to understand when in the inspiral the waveforms diverge.

We consider two families of \ac{PN} approximants for binaries where the spin
of the black hole is aligned with the orbital angular momentum:
TaylorT2~\cite{Blanchet:1996pi, Droz:1999qx, Blanchet:2006zz} and
TaylorT4~\cite{Buonanno:2002fy}.  In these models, we include all the
completely known orbital evolution terms (up to 3.5\ac{PN} order)~\cite{Wiseman:1993aj,Blanchet:1995fg,Blanchet:1995ez,Blanchet:1996pi,
Blanchet:2001ax, Blanchet:2004ek} and all the
completely known spin-related terms (up to 2.5\ac{PN}
order)~\cite{Faye:2006gx, Blanchet:2006gy, Kidder:1992fr, Mikoczi:2005dn,
Racine:2008kj}.  Restricting to systems where the spin angular momenta are
aligned (or anti-aligned) with the orbital angular momentum means that the
plane of the binary does not precess, simplifying our comparisons. However,
this study captures the dominant effect of spin on the
waveforms~\cite{Brown:2012gs}. In a separate paper, we investigate the effect
of precession on detection searches~\cite{Harry:2013effectualness}.
We also consider the
effective-one-body model as described in Ref.~\cite{Taracchini:2012ig}. 
We separately consider models that include
spin-related terms up to 3.5\ac{PN} order~\cite{Bohe:2013cla,
Blanchet:2011zv}. Spin-orbit tail (3.0\ac{PN}) and next-to-next-to-leading
order spin-orbit (3.5\ac{PN}) contributions to the phasing are known.
However, these orders are incomplete as there are also unknown spin
corrections at 3.0\ac{PN} and 3.5\ac{PN}, including spin-spin and
(spin-induced) octupole-monopole couplings. 

We restrict to comparing the inspiral portion of approximants.
Numerically modelling the merger of a black hole and a neutron 
star is an active area of research~\cite{Duez:2009yz,Shibata:2011jka,Pannarale:2012ux,Lackey:2013axa,
Foucart:2013psa}. However, producing long simulations of \ac{NSBH} systems with high spin remains a challenge, and
there is currently no widely available waveform  model that includes the complete evolution of a \ac{NSBH} coalescence 
over the full parameter space we consider. 
Refs.~\cite{Buonanno:2009zt} and ~\cite{Brown:2012nn} suggest that for non-spinning systems, 
inspiral-only templates are suitable for detection purposes below a total mass of $12\Msun$.
For a canonical 1.4 $\Msun$ neutron star, this roughly corresponds to a mass ratio of 8. 
Even at the upper range of masses we consider, $(3+15)M_{\odot}$, it has been shown in the 
case of non-spinning numerically modelled binary black hole waveforms that inspiral-only
template banks recover $> 95\%$ of the signal power~\cite{Brown:2012nn, Smith:2013mfa}. 

\begin{figure}
\includegraphics{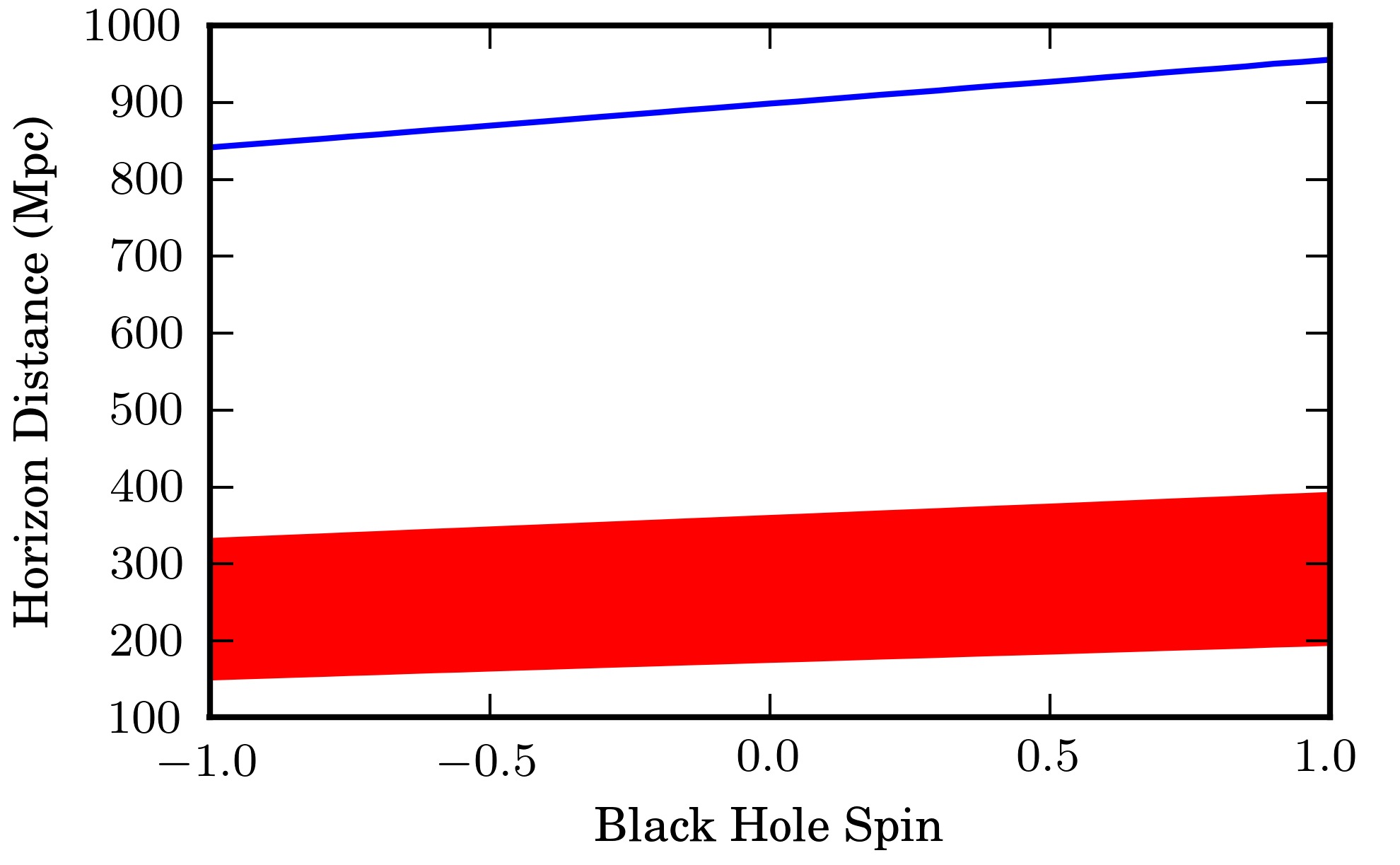}
\caption{\label{fig:t4horizon} 
The horizon distance as a function of the spin of the black hole 
for a $1.4\Msun-10\Msun$ \ac{NSBH} system, for both the \ac{aLIGO} zero-detuned,
high-power aLIGO sensitivity curve (blue) and plausible early \ac{aLIGO}
detector sensitivities (red), with a 15 Hz lower frequency cutoff. 
Results are obtained using the TaylorT4 approximant including only the 
complete spin terms up to 2.5\ac{PN}. Note that \ac{aLIGO} will be sensitive to
\ac{NSBH} systems out to $\sim 900$ Mpc, and there will be increased sensitivity
for systems with aligned black hole spins with large magnitudes. 
}
\end{figure}

In Fig.~\ref{fig:t4horizon} we show  
the distance an optimally oriented system would be observed at \ac{SNR} 8 (the horizon distance),
for a $1.4\Msun-10\Msun$ \ac{NSBH} system, as a function of the spin of the
black hole, for both the \ac{aLIGO} zero-detuned, high-power sensitivity curve and a
plausible range of early \ac{aLIGO} sensitivities~\cite{Aasi:2013wya}. 
Systems where the spin of the black hole is large in magnitude and aligned with the orbital 
angular momentum can be seen from a greater distance than systems where the spin is 
small or anti-aligned. Achieving this sensitivity requires \ac{NSBH} waveforms that do not
incur a significant loss in \ac{SNR} when used as search templates ~\cite{Apostolatos:1996rf}.
Furthermore, extracting the physics from observed signals requires faithful templates for parameter measurement.

We find that no presently available waveform model is sufficiently accurate for use
in parameter measurement. Our key results, Figs.~\ref{fig:f2f4f}-\ref{fig:seobnrf},
show the match between the various waveform families considered here.
There is a significant disagreement between the \ac{PN} approximants we
have examined, even at at low ($\chi \sim 0.4$) spins and small ($m_{BH}/m_{NS} \sim 4$) mass ratios for TaylorF2 and TaylorT4.
The match decreases as these increase with matches as low as $\sim 0.1$ observed. This motivates
the need to compute higher order \ac{PN} spin corrections. 

Our present knowledge of \ac{NSBH} waveforms will limit the
ability of gravitational-wave observatories to accurately determine source
parameters from the detected signals and may hinder the detection of some sources.
Further analytical and numerical modeling of \ac{NSBH} systems will be needed
before \ac{aLIGO} comes online in 2015 and reaches full sensitivity in $\sim$
2019~\cite{Aasi:2013wya}.

The remainder of this paper is organized as follows.  In
Sec.~\ref{sec:waveforms}, we describe the construction of the \ac{PN}
approximants used and Sec.~\ref{sec:faithfulness_definition} describes our
method of comparing them.  In Sec.~\ref{sec:faithfulness} we show the results
of comparing different \ac{PN} approximants, and show that there is a large
discrepancy between the waveforms for \ac{NSBH} binaries at relatively low
black hole spins. In Sec.~\ref{sec:R2F4} we construct a new frequency domain
approximant that is designed to agree with TaylorT4. This is followed by a
comparison of the time domain approximants to their frequency domain
counterparts in Sec.~\ref{sec:freq_vs_time_approx}, where we demonstrate that
they largely agree. Finally, in Sec.~\ref{sec:faithfulness_phase} and
Sec.~\ref{sec:faithfulness_match_accumulation} we investigate where in the
inspiral the disagreement between the waveform families becomes important. We
demonstrate that the divergence occurs at surprisingly low velocities for even
modest black hole spins. Finally in Sec.~\ref{sec:effectualness_and_flow} we
investigate whether maximizing over the mass and spin parameters of the
waveform can improve the agreement between present models, and investigate the accuracy of the
waveforms for early aLIGO observations when the detectors will have reduced
low-frequency sensitivity when compared to the ultimate sensitivity. 


\section{Constructing post-Newtonian Waveforms}
\label{sec:waveforms}

We examine the accuracy and convergence of currently known waveforms for
\ac{NSBH} binaries by comparing approximants constructed
using the \ac{PN} approximations of the binary's equation of motion and
gravitational radiation.  To obtain the gravitational-wave phase from these
quantities, we assume that the binary evolves adiabatically through a series
of quasi-circular orbits. This is a reasonable approximation as gravitational
radiation is expected to circularize the orbits of isolated
binaries~\cite{Peters:1964zz}.  In this limit, the equations of motion reduce
to series expansions of the center-of-mass energy $E(v)$ and gravitational-wave
flux $\mathcal{F}(v)$, which are expanded as a power series in the orbital
velocity $v$ around $v = 0$. They are given as
\begin{align}
E(v) &= E_{\mathrm{N}} v^2 \left(1+\sum_{n=2}^{6}E_i v^i\right), \\
F(v) &= F_{\mathrm{N}} v^{10} \left(1+\sum_{n=2}^{7}F_i v^i\right),
\end{align}
where the coefficients $\{E_\mathrm{N}, E_i, F_\mathrm{N}, F_i\}$ are
defined in Appendix~\ref{app:EF}.  For terms not involving the spin of the
objects, the energy is known to order $v^6$, while the flux is known to $v^7$,
referred to as $3.0$PN and $3.5$PN, respectively.  At order $3.0$PN, the flux
contains terms proportional to both $v^6$ and $v^6 \log v$; which are regarded
to be of the same order. Complete terms involving the spins of the objects first
appear as spin-orbit couplings at 1.5\ac{PN} order, with spin-spin couplings
entering at 2\ac{PN} order, and next-to-leading order spin-orbit couplings
known at 2.5\ac{PN} order. 

We use the assumption that these systems are evolving independently to relate
the \ac{PN} energy and gravitational-wave flux equations, i.e. the loss of energy of the
system is given by the gravitational-wave flux
\begin{equation}
\frac{dE}{dt} = - \mathcal{F}.
\end{equation}
This can be re-arranged to give an expression for the time evolution of the
orbital velocity,
\begin{equation}\label{eq:dvdt}
\frac{d v}{dt} = - \frac{\mathcal{F}(v)}{E'(v)},
\end{equation}
where $E'(v) = dE/dv$. The orbital evolution can be transformed to the gravitational
waveform by matching the near-zone gravitational potentials to the wave
zone. The amplitude of gravitational waves approximated in this way are given by the
\ac{PN} expansion of the amplitude. This gives different amplitudes for
different modes of the orbital frequency. The dominant gravitational-wave frequency $f$ is
given by twice the orbital frequency, which is related to the orbital velocity
by $v = (\pi M f)^{1/3}$. The orbital phase is therefore given by 
\begin{equation}\label{eq:dphidt}
\frac{d\phi}{dt} = \frac{v^3}{M},
\end{equation}
and the phase of the dominant gravitational-wave mode is twice the orbital phase.
Here, we will only expand the  gravitational-wave  amplitude to Newtonian
order (0\ac{PN}), which, when combined with the phase, is referred to as a
restricted \ac{PN} waveform.

Solutions $v(t)$ and $\phi(t)$ to Eqs.~(\ref{eq:dvdt}) and (\ref{eq:dphidt})
can be used to construct the plus and cross polarizations and the observed
gravitational waveform.  For restricted waveforms, these are:
\begin{eqnarray}
h_+(t) &=& - \frac{2\,M\,\eta}{D_L}\,v^2\,(1 + \cos^2 \theta)\,\cos 2 \phi(t)\
, \\
h_\times(t) &=& - \frac{2\,M\,\eta}{D_L}\,v^2\,2 \cos \theta\,\sin 2 \phi(t)\ ,
\\
h(t) &=& F_+ \, h_+(t) + F_\times \, h_\times(t)\ .
\end{eqnarray}
Here $F_+$ and $F_\times$ are the antenna pattern functions of the detector,
$D_L$ is the luminosity distance between the binary and observer, and $\theta$
is the inclination angle between the orbital angular momentum of the binary and
the direction of gravitational-wave propagation: $\cos \theta = \hat{L} \cdot \hat{N}$. Thus, a
non-precessing, restricted PN waveform is fully specified by 
$v(t)$ and $\phi(t)$ (or equivalently $t(v)$ and $\phi(v)$).

We now have the ingredients necessary to produce the TaylorT2 and TaylorT4
families of approximants, which we describe in the following sections. 

\subsection{TaylorT4}

The TaylorT4 approximant, introduced in~\cite{Buonanno:2002fy}, is formed by
numerically solving the differential equation
\begin{equation}\label{eq:t4}
\frac{dv}{dt} = \left[ \frac{-\mathcal{F}(v)}{E'(v)} \right]_{k} = A_{k}(v).
\end{equation}
The notation $\left[ Q \right]_{k}$ indicates that the quantity $Q$ is to be
truncated at $v^k$ order. Terms containing pieces logarithmic in
$v$ are considered to contribute at the order given by the non-logarithmic
part. Thus waveforms expanded to 3.5\ac{PN} order in the phase would be
truncated at $k = 7$.  We use $A_k$ as shorthand for the truncated quantity that
is used as the expression for $dv/dt$.

The resulting differential equation, given explicitly in Appendix~\ref{app:T4},
is non-linear and therefore must be solved numerically. The result is a
function $v(t)$. The phase can then be calculated by
\begin{equation}
\frac{d\phi}{dt} = \frac{v(t)^3}{M}.
\end{equation}
The phase is integrated from a fiducial starting frequency up to the \ac{MECO},
which is defined by
\begin{equation}
\frac{dE(v)}{dv} = 0.
\end{equation}
The \ac{MECO} frequency is where we consider the adiabatic approximation to have broken down. Note
that the \ac{MECO} frequency is dependent on not only the masses but also the
spins of the objects; specifically, systems where the objects' spins are aligned
with the orbital angular momentum will have a higher \ac{MECO} frequecy.
When the partial spin-related terms at 3.0\ac{PN} and 3.5\ac{PN} are included, however,
there are regions of the \ac{NSBH} parameter space for which the MECO condition is never satisfied. For these cases,
we impose that the rate of increase in frequency must not decrease (i.e. we stop if $dv/dt \leq 0$), and that the 
characteristic velocity of the binary is less than $c$ (i.e. we stop if $v \geq 1$). 
We terminate the waveforms as soon as any of these stopping conditions are met.


\subsection{TaylorT2}
\label{subsec:t2}

In contrast to the TaylorT4 approximant, the TaylorT2 approximant is
constructed by expanding $t$ in terms of $v$ and truncating the expression to
consistent \ac{PN} order. We first construct the quantity
\begin{equation}\label{eq:t2}
\frac{dt}{dv} = \left[ \frac{E'(v)}{-\mathcal{F}(v)} \right]_{k} = B_{k}(v).
\end{equation}
This can be combined with the integral of~\eqref{eq:dphidt} and solved in
closed form as a perturbative expansion in $v$,
\begin{equation}\label{eq:phaset2}
\phi(v) = \int \frac{v^3}{M} B_{k}(v) dv.
\end{equation}
The explicit result of this integral is given in Appendix~\ref{app:T2}.
Similar to TaylorT4, the phase is generally calculated up to the \ac{MECO}
frequency. However, for some points of parameter space, this formulation can
result in a frequency that is not monotonic below the \ac{MECO} frequency.
As with TaylorT4, we stop the waveform evolution with $dv/dt \leq 0$ or $v \geq 1$.

A related approximant can be computed directly in the frequency domain by using
the stationary phase approximation~\cite{Droz:1999qx,Blanchet:2006zz}.  This
approximant is called TaylorF2 and can be expressed as an analytic expression
of the form
\begin{equation}
\phi(f) = A(f) e^{ i \psi(f) },
\end{equation}
where the phase takes the form
\begin{equation}
\psi(f) = \sum_{i = 0}^{7} \sum_{j = 0}^{1} \lambda_{i, j} f^{(i-5)/3} \log^j
f.
\end{equation}
The full expressions for the amplitude and phase are given in
Appendix~\ref{app:F2}.  Because the stationary phase approximation is generally
valid, the TaylorT2 and TaylorF2 approximants are nearly
indistinguishable~\cite{Droz:1999qx}. An advantage of the TaylorF2 approximant
comes from the fact that it can be analytically calculated in the frequency
domain.  In practice, waveforms that are generated in the frequency domain
without the use of integration are less computationally costly, and so searches
for  gravitational waves from inspiraling binary systems have been performed using the
TaylorF2 approximant~\cite{Blanchet:1996pi, Droz:1999qx, Blanchet:2006zz,
Abbott:2003pj, Abbott:2005pf, Abbott:2005pe, Abbott:2005qm, Abbott:2007xi,
Abbott:2009zi, Abbott:2009qj, Abadie:2010yba, Abadie:2011nz, Abbott:2007rh,
Abadie:2010uf, Briggs:2012ce}.

%

\subsection{SEOBNRv1}

An additional approximant we use is the spinning effective one-body model
(SEOBNRv1), presented in Ref.~\cite{Taracchini:2012ig}.  This approximant
incorporates the results of black hole perturbation theory, the self-force
formalism and \ac{PN} results. The model has been calibrated to numerical
relativity simulations, including simulations where the objects' spins were
(anti-) aligned with the orbital angular momentum and had magnitudes of $\chi
\pm 0.4$.  In order to compare these waveforms more fairly with the \ac{PN}
approximants that only model the inspiral, we truncate this model before the
merger section of the waveform.

The implemented versions of SEOBNRv1 are currently limited to $\chi \leq 0.6$.
To further extend the model would require better modeling of the plunge physics
and possibly the computation and incorporation of additional \ac{PN} terms.

\section{Computing faithfulness}
\label{sec:faithfulness_definition}

Searches for gravitational waves from compact binary coalescences utilize
matched-filtering~\cite{Wainstein,Allen:2005fk}, in which the signal model is
correlated with the detector output to construct a signal-to-noise
ratio. If the signal model does not accurately capture the true gravitational
waveform, then the signal-to-noise ratio, and hence the distance to which the
detector can see signals at a given false alarm rate, will decrease. Matched-filtering 
therefore relies on the accuracy of the models. We quantify the 
agreement between waveform families by computing the match, or
\emph{faithfulness} of the waveforms, defined as follows.
We 
define the noise-weighted inner product between two gravitational waveforms, $h_1$ and
$h_2$, to be
\begin{equation}
(h_1|h_2) = 4 \Re{
\int_{0}^{\infty}\dfrac{\tilde{h}_1(f)\tilde{h}_2^*(f)}{S_n(f)} df},
\end{equation}
where 
\begin{equation}
\tilde{h}_1(f) = \int_0^\infty h_1(t) e^{-2\pi i ft} \, dt
\end{equation}
is the Fourier transform of $h_1(t)$, and $S_n(f)$ denotes the one-sided power spectral
density of the gravitational-wave detector's noise. In practice, the signals are discretely sampled so the upper
frequency limit is the Nyquist frequency of the data,
and the lower frequency limit of the integral is set by the detector's
low-frequency sensitivity~\cite{Allen:2005fk}. We define the normalized overlap between two
waveforms $h_1$ and $h_2$ as
\begin{equation}
(h_1|h_2) = \dfrac{(h_1|h_2)}{\sqrt{(h_1|h_1)(h_2|h_2)}}.
\end{equation}
The match between two waveforms is obtained by maximizing the overlap over the phase of the waveform and $\phi_c$ and any time shifts $t_c$ between
$h_1$ and $h_2$
\begin{equation}\label{eq:match}
\mathcal{M}(h_1,h_2) = \underset{\phi_c,t_c}{\max} (h_1|h_2(\phi_c,t_c)),
\end{equation}
where the shifted waveform can be constructed as
\begin{equation}
\tilde{h}(\phi_c,t_c) = \tilde{h} e^{i\left(\phi_c - 2 \pi f t_c \right)}.
\end{equation}
The \emph{faithfulness} of representing a waveform from a given \ac{PN} family with
that of another is described by the match between the two waveforms when the
same physical parameters are used as input to the models. As both models
describe the same physical source, the match should be unity. Any deviation is
due to the variation between models and the match gives the fractional loss in
signal-to-noise ratio that will result.

\section{Post-Newtonian approximant faithfulness comparison}
\label{sec:faithfulness}

\begin{figure}
\includegraphics{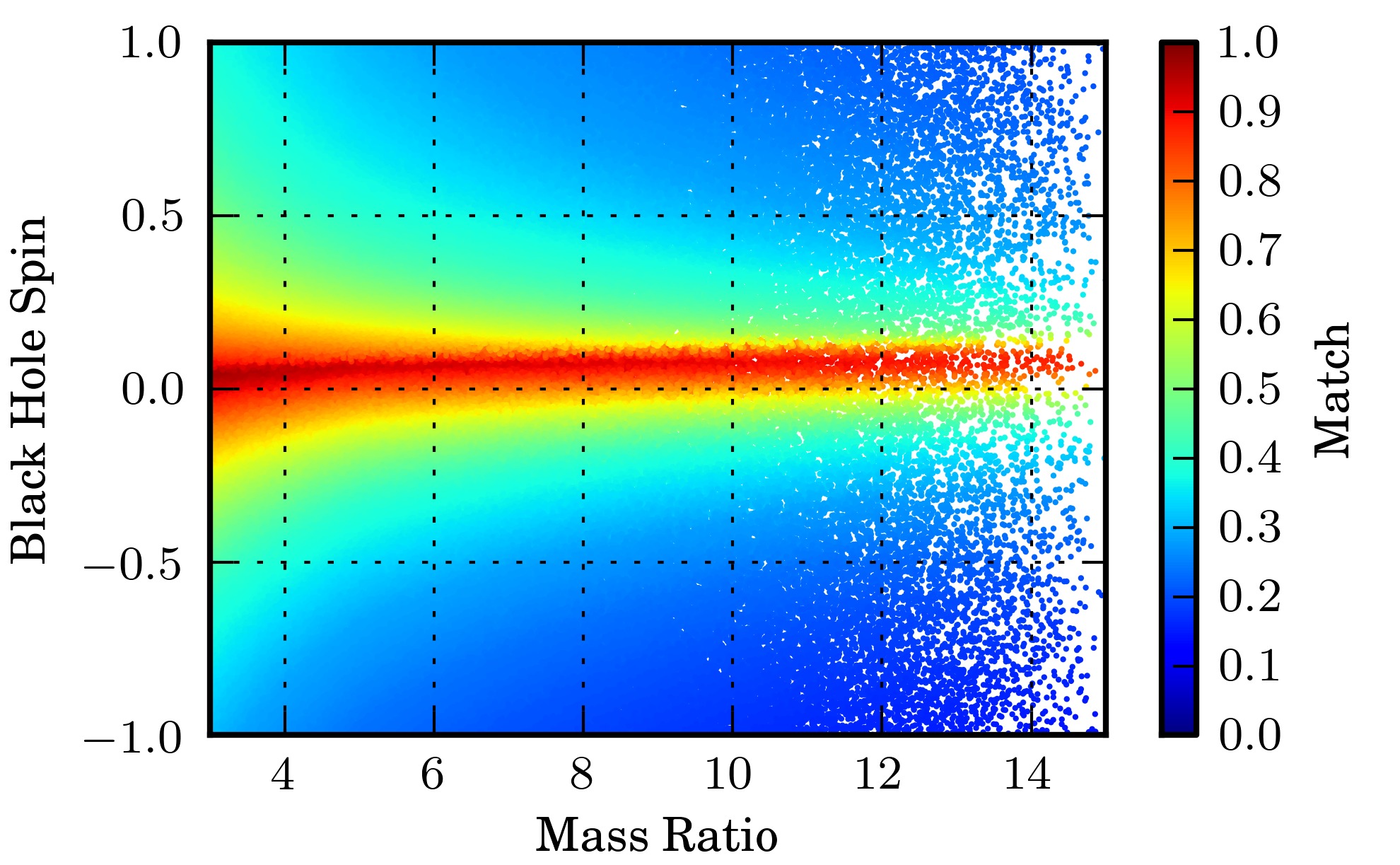}

\caption{\label{fig:f2f4f}The match between the TaylorF2 and
TaylorT4 approximants as a function of the spin of the black hole
and the mass ratio of the system. Only the completely known 
spin-related corrections up to 2.5\ac{PN} are included. Matches are calculated using the
the aLIGO zero-detuned, high-power sensitivity curve and a 15Hz lower frequency cutoff.
A significant reduction in match is seen for even moderate spins $\chi \sim 0.3$
and low mass ratios $m_{bh}/m_{ns} \sim 4$. The approximants also begin to disagree for non-spinning
systems as the mass ratio increases.
}

\end{figure}

\begin{figure}
\includegraphics{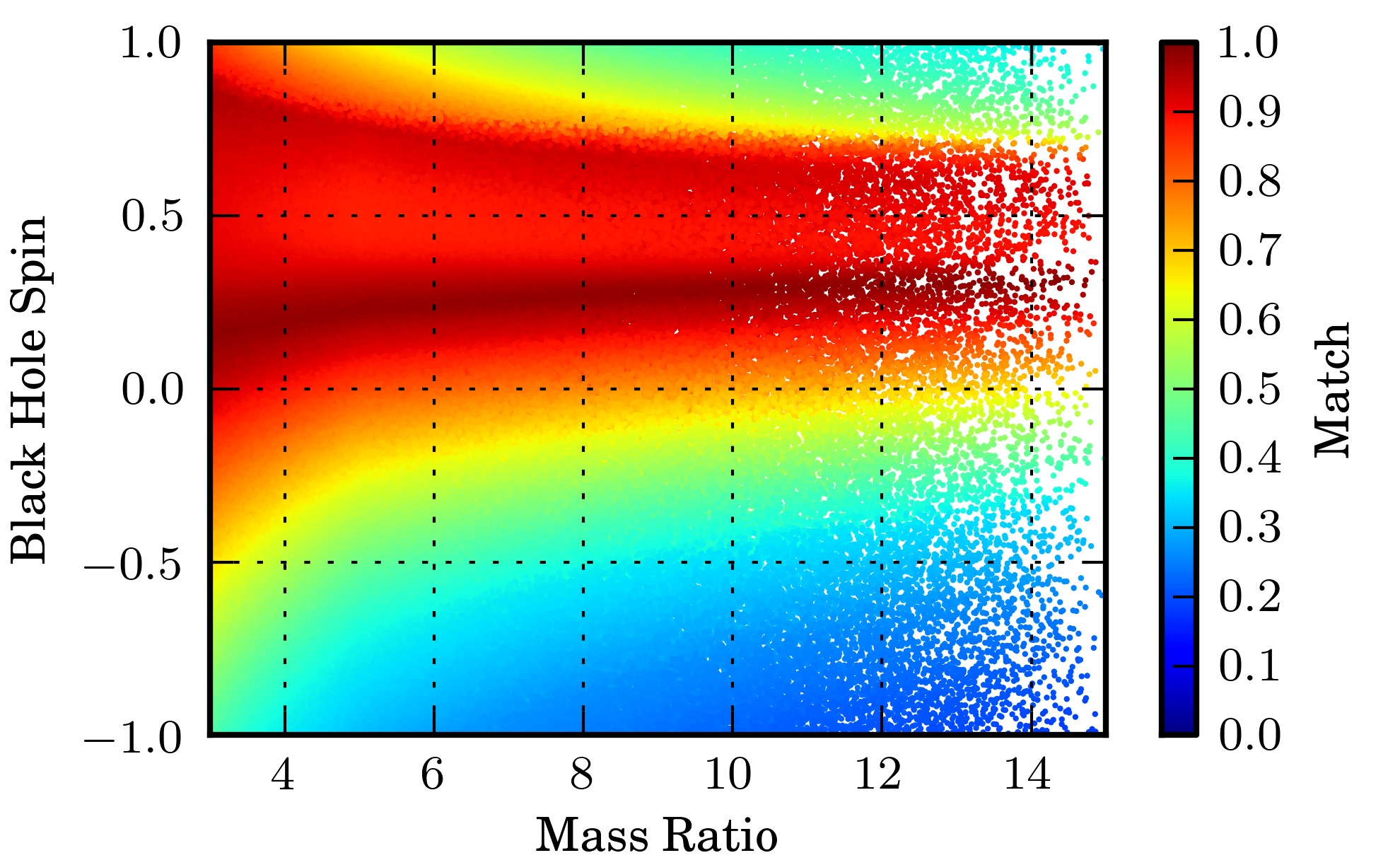}

\caption{\label{fig:f2t4fso}The match between the TaylorF2 and TaylorT4 approximants
as a function of black hole spin and mass ratio. Both models include
the next-to-next-to-leading spin-orbit (3.5\ac{PN}) and spin-orbit tail 
terms (3.0\ac{PN}). In comparison to Fig. \ref{fig:f2f4f}, the additional terms have 
improved the agreement for moderately spinning aligned spin systems, however, the
match is still $ \sim 0.8 $ for $\chi \sim 0.5 $ at all mass ratios.}

\end{figure}

In this section we compare the faithfulness between waveforms from different
\ac{PN} approximants where we choose the physical parameters to be consistent
with \ac{NSBH} sources.  We also consider how the waveforms from the \ac{PN}
approximants compare to the waveforms from the SEOBNRv1 effective-one-body
model~\cite{Taracchini:2012ig}. Lastly, we consider the effect of including 
the spin-related terms at only partially derived orders. 
We model the sensitivity of second generation  gravitational-wave detectors with the aLIGO
zero-detuned, high-power sensitivity curve~\cite{aLIGOSensCurves}. For this
study we use a lower frequency cutoff of 15Hz since it is not expected that
detectors will have significant sensitivity below this frequency. We consider
the effect of increasing this low-frequency cutoff to simulate early aLIGO
sensitivities in Sec.~\ref{sec:effectualness_and_flow}.

\begin{figure}
\includegraphics{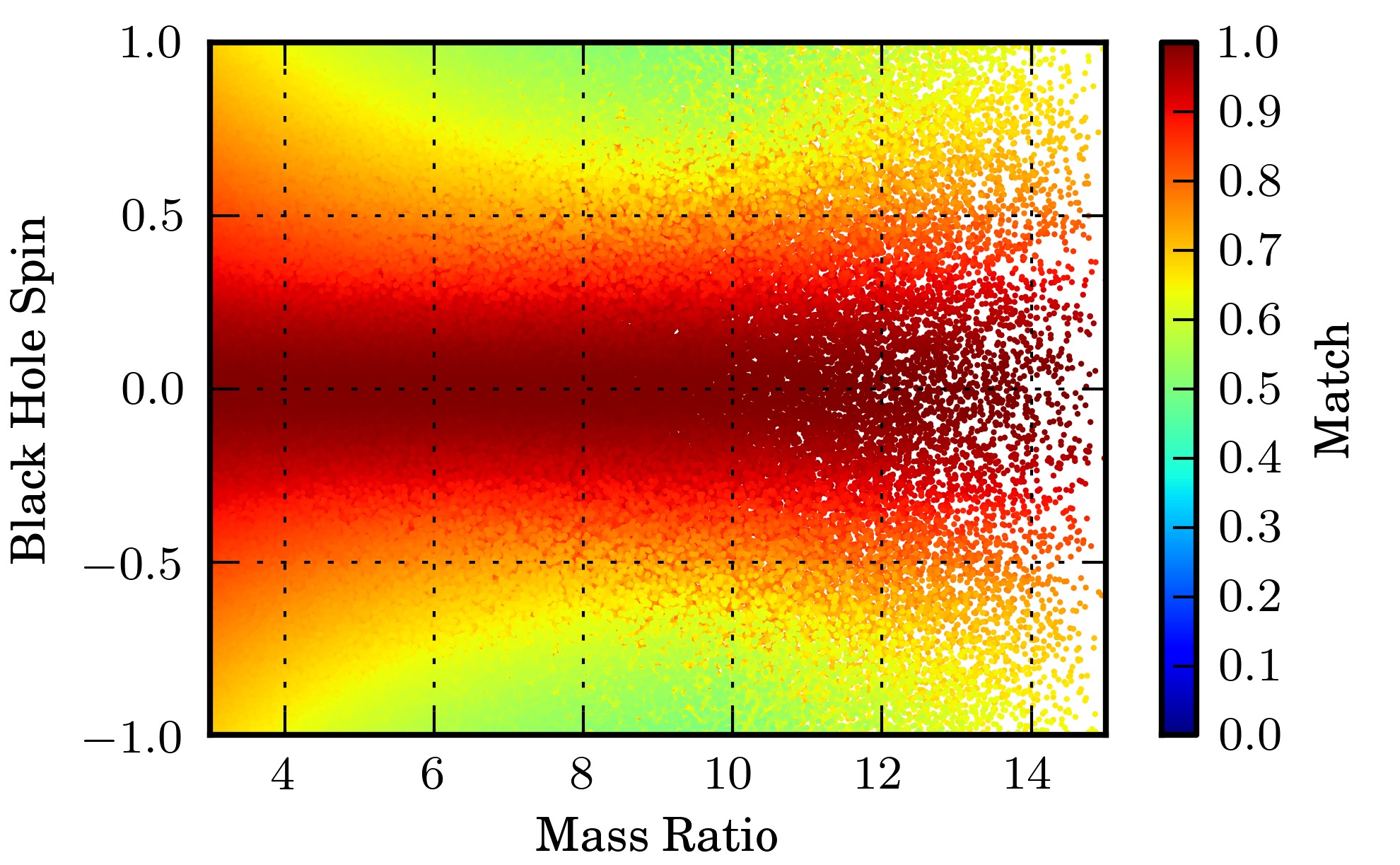}
\caption{\label{fig:5715f2f2}The match between TaylorF2 with 2.5\ac{PN} spin corrections
and TaylorF2 including the next-to-next-to-leading spin-orbit (3.5\ac{PN}) and spin-orbit tail 
terms (3.0\ac{PN}), as a function of the spin of the black hole
and the mass ratio of the system. Matches are calculated using the the aLIGO
zero-detuned, high-power sensitivity curve and a 15Hz lower frequency cutoff. Although 
there is agreement where the spins are low $\chi < 0.2 $, the match quickly drops
as the spin of the black hole increases, so that the match is already $ \sim 0.7 $ for $\chi \sim 0.5$.
}
\end{figure}

\begin{figure}
\includegraphics{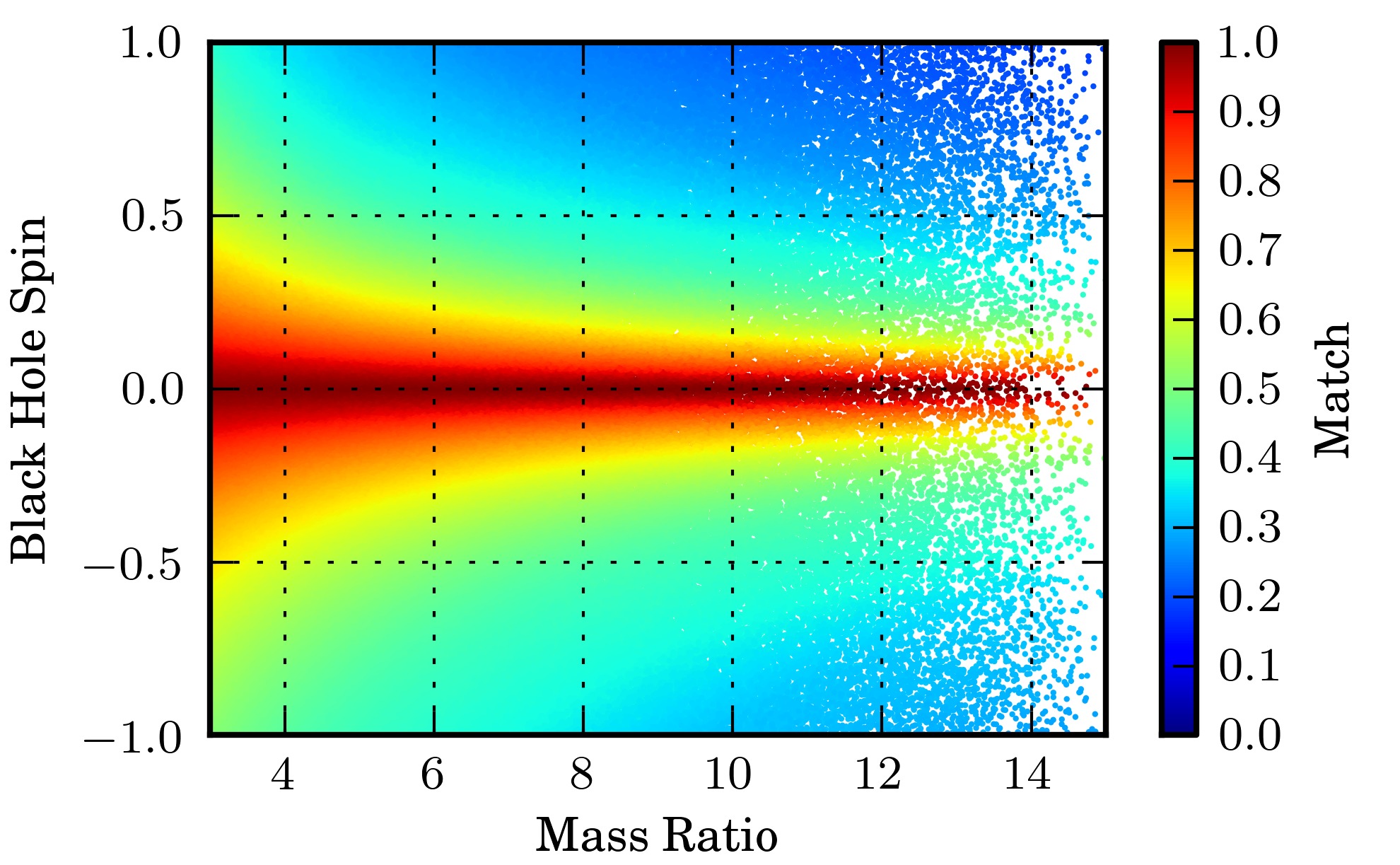}
\caption{\label{fig:5715t4t4}The match between TaylorT4  with 2.5\ac{PN} spin corrections
and TaylorT4 including the next-to-next-to-leading spin-orbit (3.5\ac{PN}) and spin-orbit tail 
terms (3.0\ac{PN}), as a function of the spin of the black hole
and the mass ratio of the system. Matches are calculated using the the aLIGO
zero-detuned, high-power sensitivity curve and a 15Hz lower frequency cutoff. 
In comparison to Fig.~\ref{fig:5715f2f2}, the approximant is more noticeably changed
by the additional terms. For a mass ratio of 8, the match has already 
fallen to  $ \sim 0.7 $ for $\chi \sim 0.15$. }
\end{figure}

In Fig.~\ref{fig:f2f4f}, we examine the faithfulness of \ac{NSBH} waveforms by computing the match between the TaylorF2 and TaylorT4 \ac{PN} approximants.
The TaylorT4 approximant was used to simulate \ac{NSBH} binaries in LIGO's
previous gravitational-wave searches, and the TaylorF2 family is used as the
templates for detection~\cite{Abadie:2011nz}.
In order to focus
on the mismatches primarily due to phase differences between the models, the
frequency cutoff of the TaylorF2 waveform is made to agree with the ending
frequency of the TaylorT4 waveform. We see that the agreement between the two
models is primarily influenced by the magnitude of the black hole's spin, and
secondarily by the mass ratio. There is a noticeable drop in match at higher
mass ratios, even when
the spin of the black hole is zero. As expected, the best
agreement is seen when the black hole's spin is small and
the black hole and neutron star have comparable masses.
However, this plot shows that there is a \emph{substantial} disagreement between
these approximants for even moderately low black hole spins ($\chi \sim 0.3$),
which increases as the spin of the black hole increases. 
We note that the effect on the match due to the spin of the
neutron star is negligible in all areas.  In Fig.~\ref{fig:f2t4fso} we compare
the TaylorF2 and TaylorT4 models, with the inclusion of the spin-orbit
tail (3.0\ac{PN}) and next-to-next-to-leading spin-orbit (3.5\ac{PN})
corrections recently computed in Refs.~\cite{Bohe:2012mr, Blanchet:2012sm}.  In comparison to
Fig.~\ref{fig:f2f4f}, the agreement is significantly improved for aligned spins
with moderate magnitudes. However, these approximants maintain a poor level of
overall agreement, with matches of only $\sim 0.8$ at $\chi \sim 0.5$ for all mass ratios, and even
lower matches for anti-aligned systems. 
Figs.~\ref{fig:5715f2f2} and~\ref{fig:5715t4t4} compare the TaylorT2 and TaylorT4 approximants with and without
these additional spin terms.
We see that TaylorT4 is especially sensitive to the additional corrections.
In both cases, however, we note that the additional terms have caused a significant change in the waveforms, 
as indicated by the low matches, demonstrating that the expansion has
not yet sufficiently converged to produce reliable waveforms for parameter estimation. 

\begin{figure*}
\centering
\begin{minipage}[l]{2.0\columnwidth}
\centering
\includegraphics{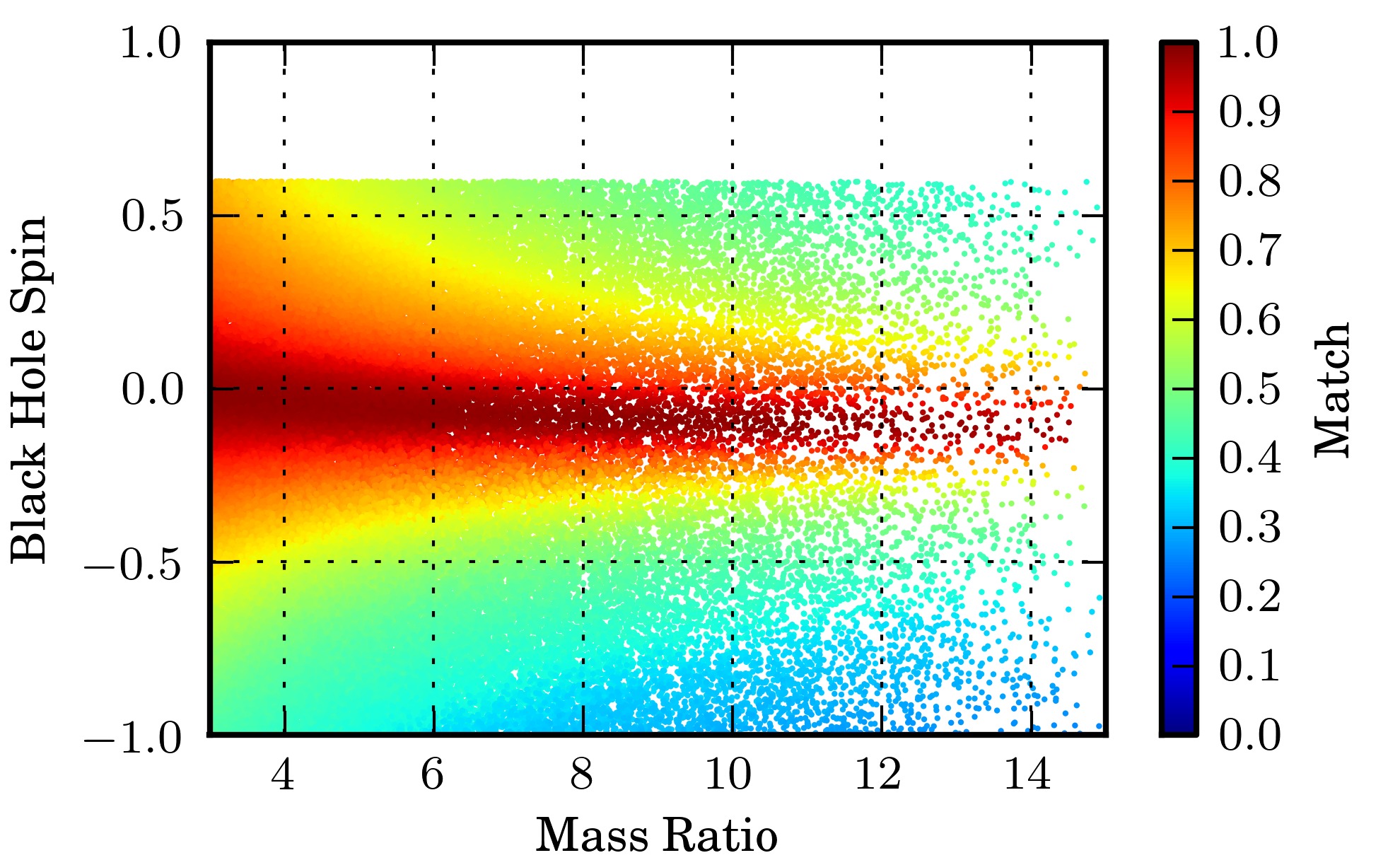}
\includegraphics{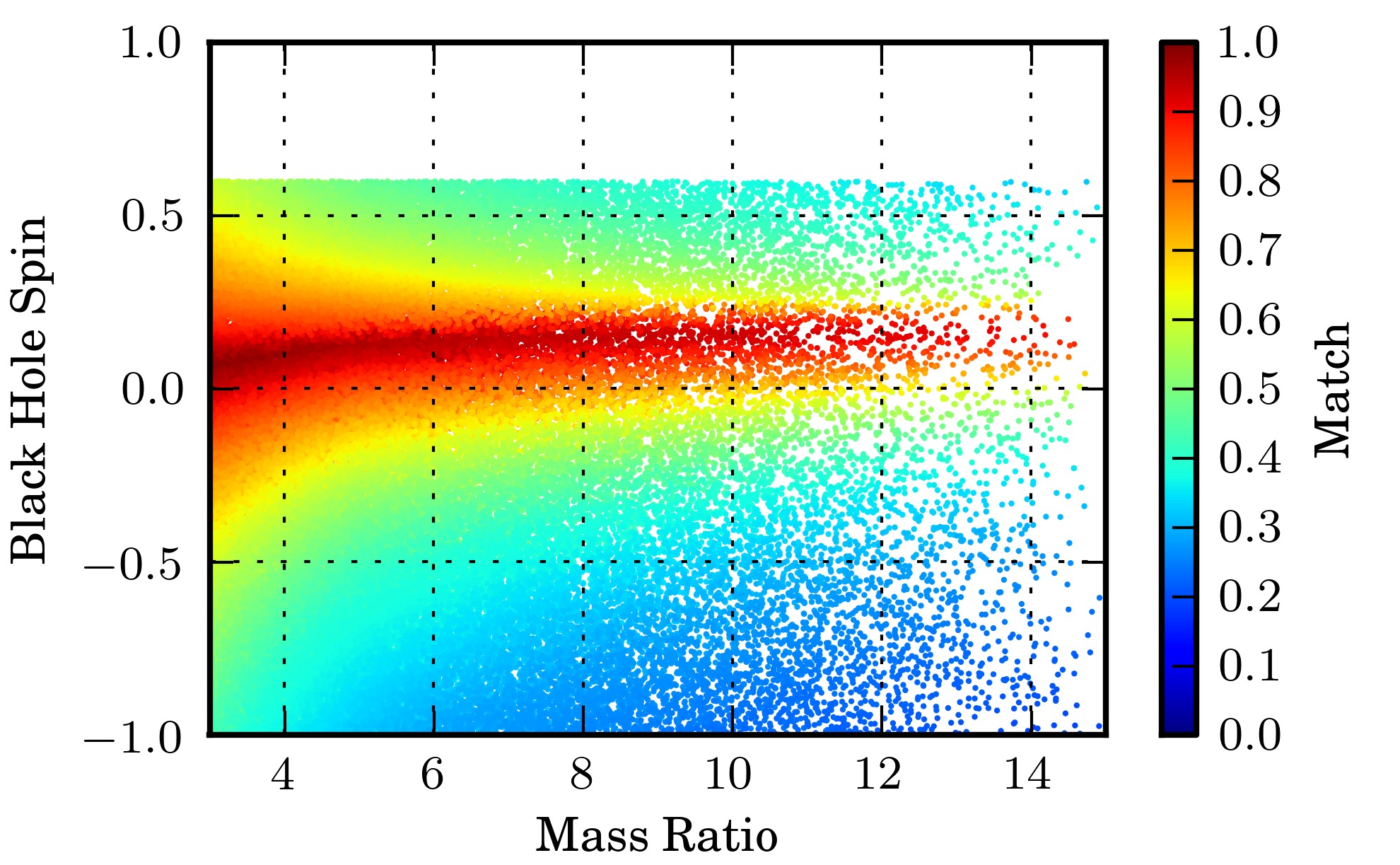}
\caption{\label{fig:seobnrf}The match between the TaylorF2
(left) or TaylorT4 (right) and SEOBNRv1
approximants. Spin corrections for the \ac{PN} approximants are included up to 2.5\ac{PN}.
Matches are calculated using the the aLIGO zero-detuned, high-power sensitivity curve
with a 15 Hz lower frequency cutoff. As in Fig.~\ref{fig:f2f4f}, 
there is a significant reduction in match where
spin of the black hole is only moderate. Note, however, that the
\ac{PN} approximants have marginally better agreement with SEOBNRv1 than with each other.}
\end{minipage}
\end{figure*}

In Fig.~\ref{fig:seobnrf} we compare the SEOBNRv1 model to the \ac{PN} models
TaylorF2 and TaylorT4.  Since the SEOBNRv1 model is not valid
for large values of $\chi$~\cite{Taracchini:2012ig} we restrict
$\chi < 0.6$ and only report matches below this limit.  We see that,
similar to the comparison between TaylorF2 and TaylorT4, these models also have
large mismatches when the spin of the black hole is nonzero.
The large discrepancy between the waveform families indicates that higher order
\ac{PN} correction terms are required. This may also pose significant problems
for parameter estimation of \ac{NSBH} sources.



\section{The TaylorR2F4 approximant}
\label{sec:R2F4}

In the previous section, we found a surprisingly large disagreement between the TaylorF2 and TaylorT4
\ac{PN} approximants when compared with waveform parameters appropriate for
\ac{NSBH} systems. We would like to distinguish how much of this is due to
differences between time domain and frequency domain approximants, and how much
of this is due to differences between the formulation of the two \ac{PN}
families.  This can easily be performed for the TaylorF2 and TaylorT2
approximants, however we need to construct an equivalent frequency domain
version of TaylorT4 to complete the comparison.

By analogy with TaylorF1 and TaylorF2~\cite{Damour:2000zb,Buonanno:2009zt},
TaylorF4 is obtained by numerically integrating the reciprocal of Eq.~\eqref{eq:t4}
in the frequency domain,
\begin{equation}\label{eq:f4}
dt/dv = 1 / A_k(v).
\end{equation}
However, this does not elucidate the differences between the TaylorT4 and
TaylorF2 approximants. Instead, we construct an analytical approximation to the
TaylorF4 approximant, which we call TaylorR2F4, by expanding Eq.~\eqref{eq:f4} in
powers of $v$. In order to make this series finite, we truncate these
additional terms at an order in $v$ higher than the order where the \ac{PN}
expansion of the energy and flux were truncated,
\begin{equation}
\frac{dt}{dv} = \left[ \frac{1}{A_{k}(v)} \right]_l = B_{k}(v) + R_{kl}(v) =
C_{kl}(v).
\end{equation}
Here $B_{k}(v)$ is the same as in the TaylorT2 approximant and $R_{kl}(v)$ are
the terms from order $v^{k+1}$ up to order $v^l$. It is important to note that
this produces a power series that is identical to the TaylorF2 approximant up
to the point where~\eqref{eq:t2} was truncated.  Thus, terms of higher order in
$v$ account for the differences between the TaylorT2 and TaylorT4 approximants.

In sec.~\ref{sec:freq_vs_time_approx} we show that TaylorR2F4 agrees well with the TaylorT4 approximant
when expanded to $v^9$ or $v^{12}$, which we shall see in the next section.  As
noted above, the second expansion in the TaylorR2F4 approximant is a different
expansion than the \ac{PN} expansion of the energy and flux.  The Fourier phase
for the TaylorR2F4 approximant can be obtained from~\eqref{eq:phaset2}  where
$B_{k}(v)$ is replaced by $C_{kl}(v)$.  This is given up to order $v^N$ as
\begin{equation}
\psi_{\mathrm{R2F4}}(f) = \psi_{F2}(f) + \sum_{i=6}^{N} \sum_{j=0}^{N}
\lambda_{i, j} f^{(i-5)/3} \log^j f,
\end{equation}
where the form of these expressions up to $N=12$ can be found in
Appendix~\ref{app:R2F4}.
Because this approximant can be analytically expressed in the frequency domain,
it can be generated relatively cheaply compared to TaylorT4. This means that it has
the potential to be used where computational efficiency and 
a higher degree of agreement with TaylorT4 is desired.
We note that the frequency-domain approximants are much faster than their
time-domain counterparts, which must integrate differential equations and perform 
a Fourier transform. Therefore, they are especially useful in computational problems 
which are waveform-generation limited, 
such as parameter estimation of signals~\cite{Aasi:2013jjl}.

\section{Comparison of Frequency to Time Domain Approximants}
\label{sec:freq_vs_time_approx}

\begin{figure}
\includegraphics{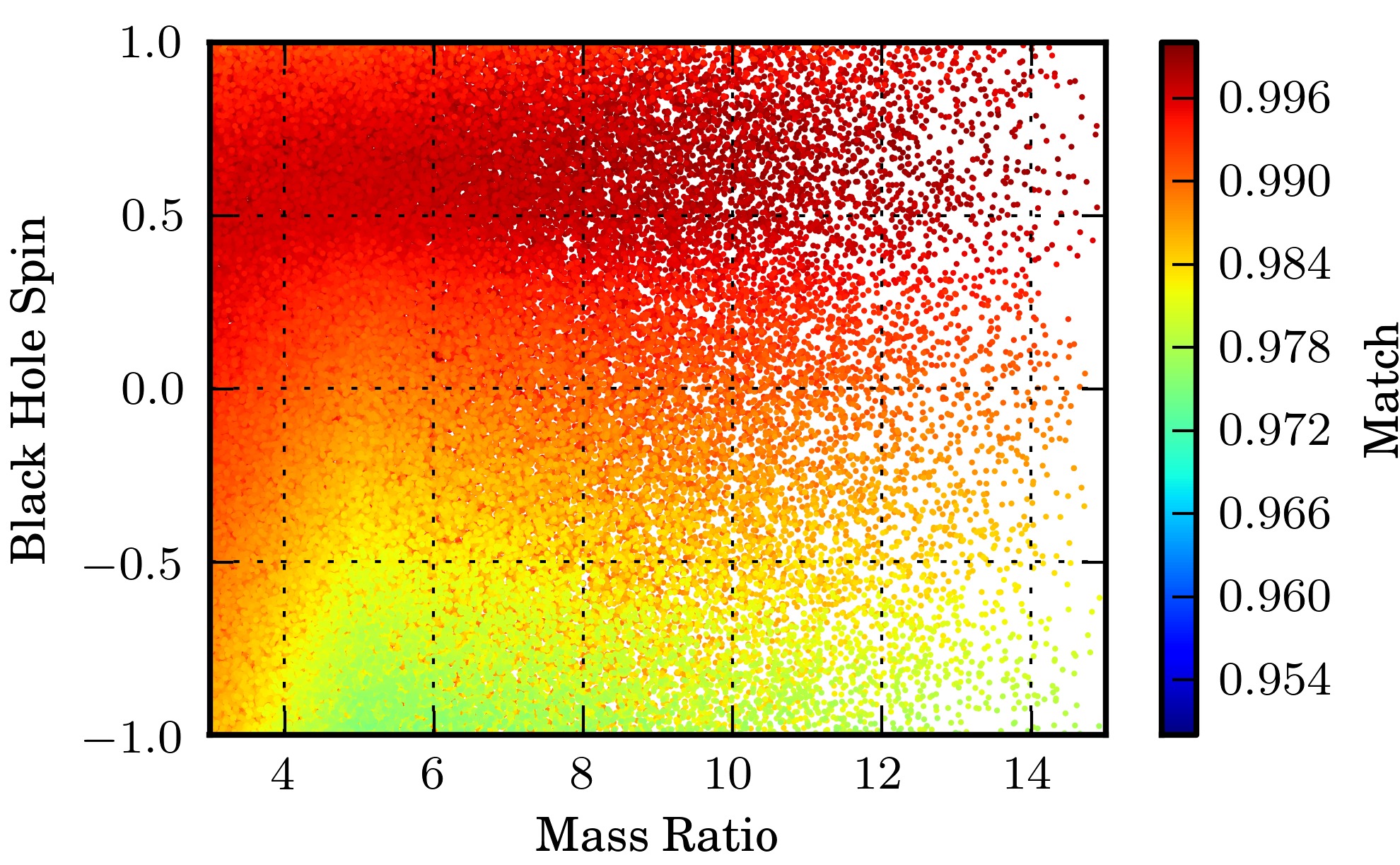}

\caption{\label{fig:f2t2fs} The match between TaylorF2 and TaylorT2. Both include spin 
corrections up to 2.5\ac{PN} order.
Matches are calculated using the the aLIGO
zero-detuned, high-power sensitivity curve and a 15Hz lower frequency cutoff. 
We see that the F2 and T2 approximants largely agree. The discrepancy
between the two approixmants can be reduced by expanding the frequency sweep of
the TaylorF2 approximant's amplitude to higher \ac{PN} orders. However, there
is different Gibbs phenomena between the two approximants that will cause a
discrepancy.}

\end{figure}

\begin{figure*}
\centering
\begin{minipage}[l]{2.0\columnwidth}
\centering
\includegraphics{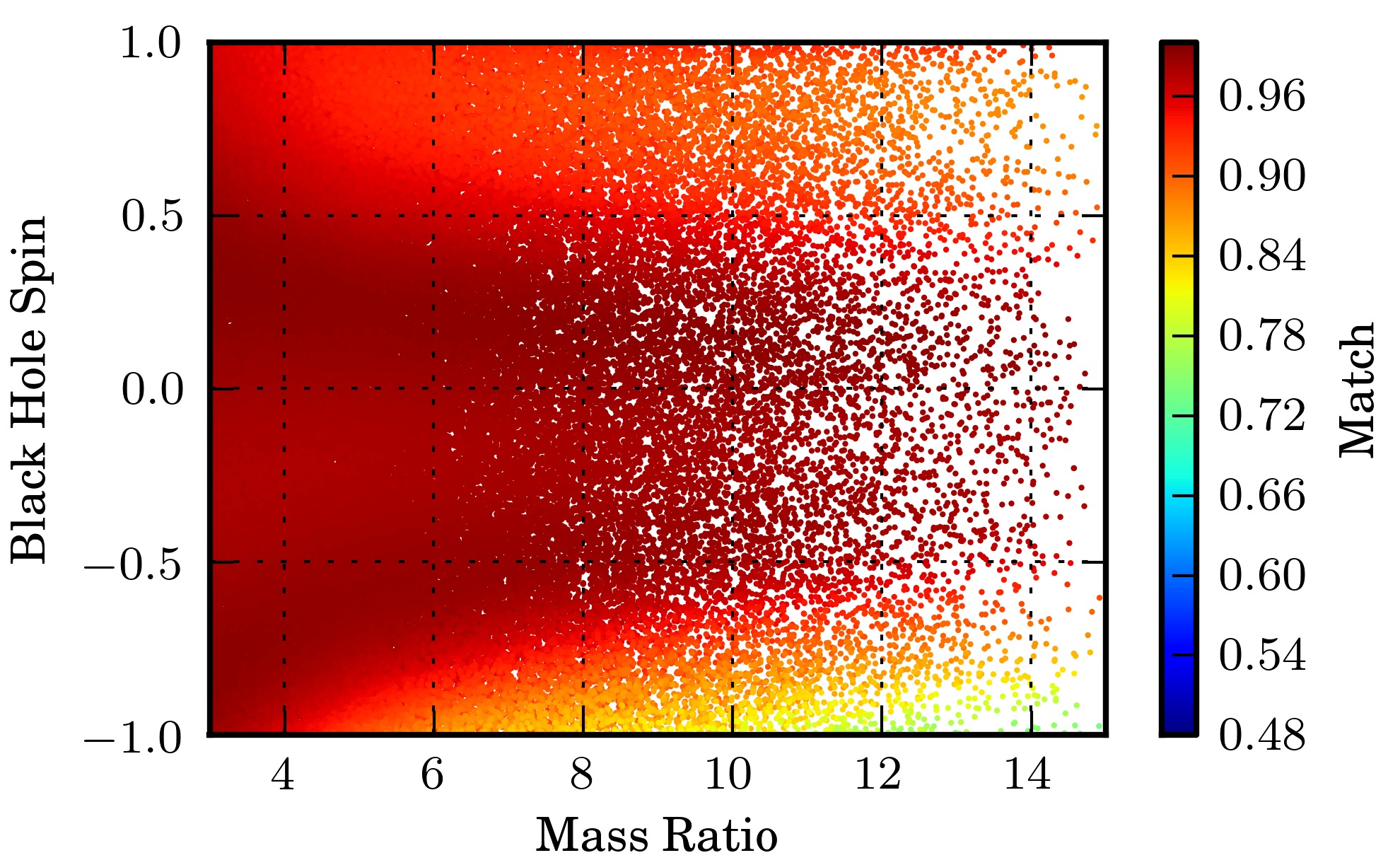}
\includegraphics{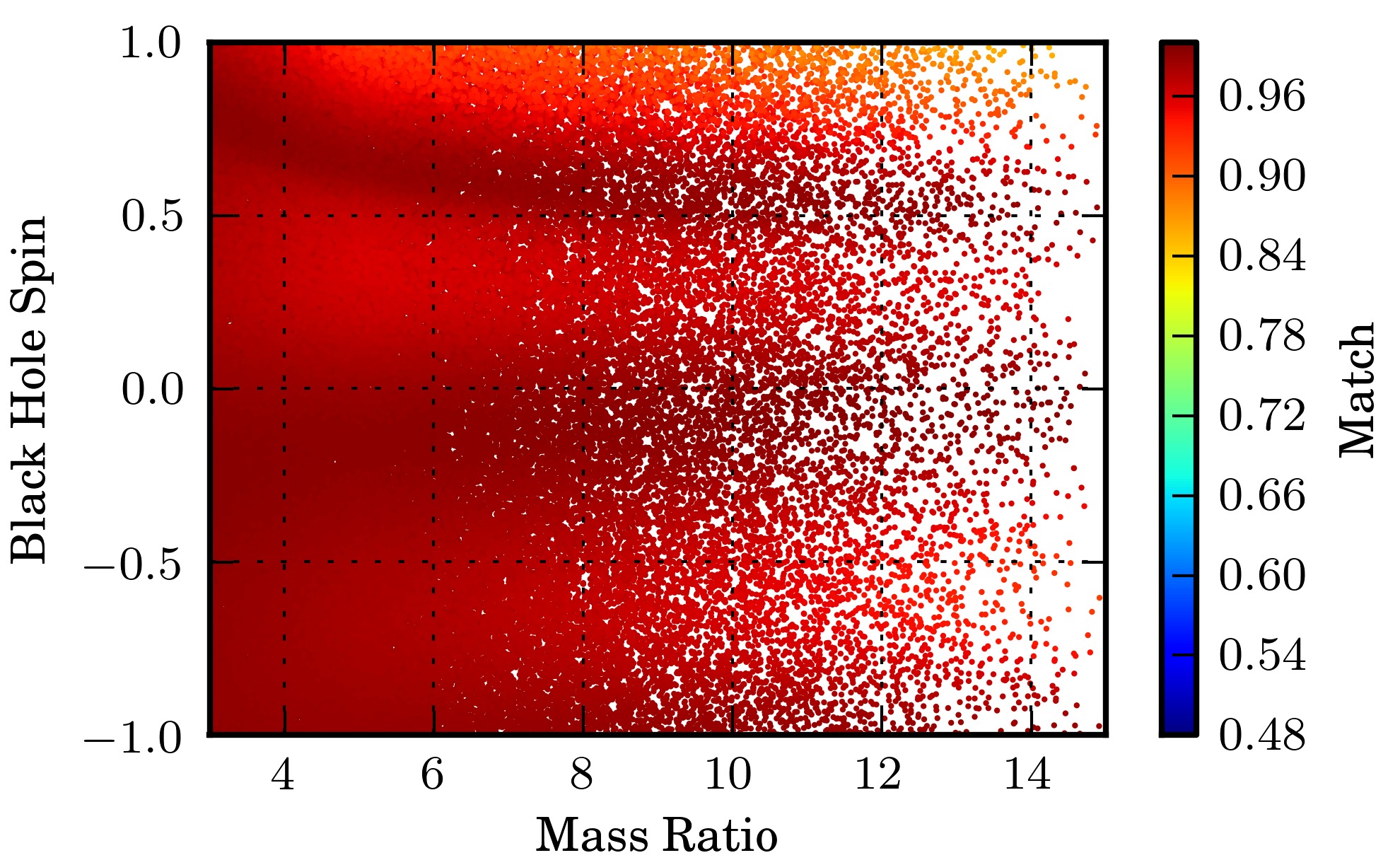}
\caption{\label{fig:f4t4fs}The match between TaylorT4 and TaylorR2F4. Both models include
spin corrections up to 2.5 \ac{PN}. TaylorR2F4 is  re-expanded up to order $v^9$ (left) 
and $v^{12}$ (right). Matches are calculated using the the aLIGO
zero-detuned, high-power sensitivity curve and a 15Hz lower frequency cutoff. R2F4 and T4 have
high agreement over a broad range of parameters, with some visible exceptions.
Expanding up to order $v^{12}$ has generally increased
agreement with TaylorT4. }
\end{minipage}
\end{figure*}
In this section, we investigate to what extent the discrepancy between the waveform families that
was demonstrated in Sec.~\ref{sec:faithfulness} is due to the difference
between expressing approximants in the frequency and time domain alone.
We compare the new TaylorR2F4 approximant from
Sec.~\ref{sec:R2F4}, and TaylorF2, to their time domain equivalents.

We find that TaylorF2 waveforms are a good representation of
TaylorT2 waveforms, even when we consider waveforms from \ac{NSBH} systems
where the component objects are spinning. This can be seen in
Figure~\ref{fig:f2t2fs}, which shows the match between the TaylorF2 and
TaylorT2 models. In that figure, the ending frequency of both models is made to
be the same, which is accomplished by terminating the TaylorF2 waveforms at the
frequency where the generation of the equivalent TaylorT2 waveforms terminated.
We find that the TaylorF2 and TaylorT2 waveforms agree to better than $\gtrsim
95.7\%$ for the entire region investigated. For systems where the black hole
spin was positively aligned with the orbital angular momentum, the match is
$\gtrsim 97.9\%$. The discrepancy between these two models is in part due to
expanding to only Newtonian order the frequency sweep associated with the
stationary phase approximation of the TaylorF2 approximant. In addition, part
of the discrepancy results from Gibbs phenomena differences between the
approximants.
It is important to note that neither of these waveforms have termination
conditions that are determined by the physical behavior of the inspiralling
binary. The termination frequency only indicates the point at which the
approximant is certainly no longer valid. The increased match for aligned spin
waveforms is due to the higher frequency cutoff, which pushes the termination
frequency out of the most sensitive part of the zero-detuned, high-power aLIGO
sensitivity curve.

Figure~\ref{fig:f4t4fs} shows a comparison between the TaylorR2F4 and TaylorT4
models. In that figure, the second expansion associated with the TaylorR2F4
model is extended to order $v^{9}$ (left) and $v^{12}$ (right), and the ending frequency of both
is that corresponding to the \ac{MECO}.  We show that the TaylorR2F4 model is
adequate for a large range of parameters as a computationally inexpensive
substitute for TaylorT4. 

Since the mismatch between the TaylorF2 and TaylorT4 models is not due to
differences between the time domain and frequency domain approximants, this
indicates that the effective higher order \ac{PN} terms used in the
construction of TaylorR2F4, which are also intrinsically present in TaylorT4,
are still significant. To obtain better agreement between the
different \ac{PN} approximants we consider, it is necessary to extend the
\ac{PN} expansions of the energy and flux equations to include 
unknown higher order terms, particularly ones that involve 
the spin of the objects. 

\section{Accumulation of Phase Discrepancy}
\label{sec:faithfulness_phase}

\begin{figure*}
\includegraphics[]{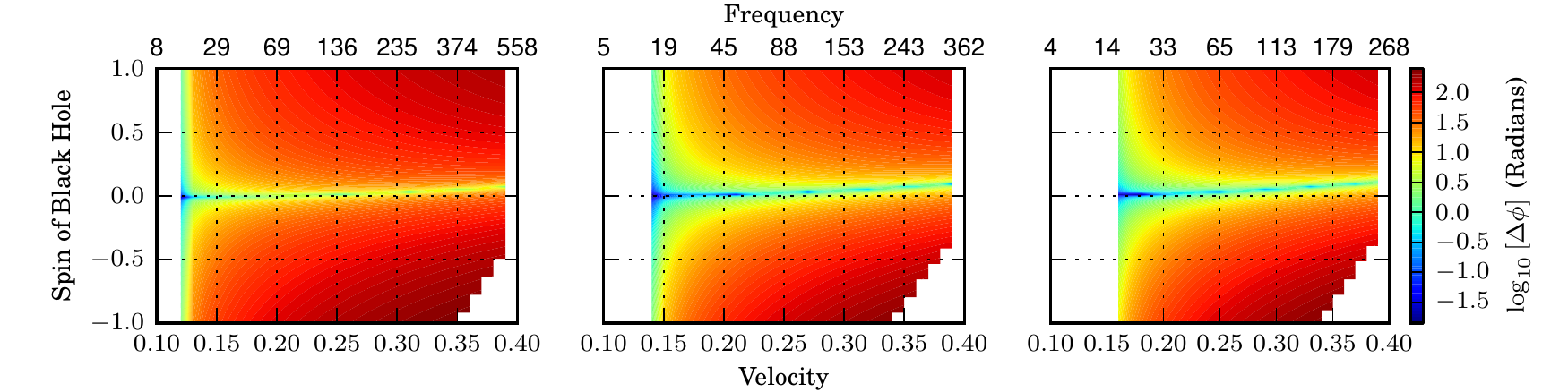}

\caption{\label{fig:T2T4pa} The accumulation of phase differences between
TaylorT2 and TaylorT4, for systems with component masses $(m_1, m_2)$ of
$(1.4\Msun, 6\Msun)$ (left), $(1.4\Msun, 10\Msun)$ (center), and $(1.4\Msun,
14\Msun)$ (right). The approximants include spin terms up to 2.5\ac{PN}.
The calculation starts from the velocity corresponding to a
 gravitational-wave  frequency of $15$Hz, continues to the velocity on the horizontal axis,
and reports the difference in accumulated gravitational-wave phase between the waveforms. The feature
in the bottom right corner of each plot arises because the TaylorT2 approximant is no longer monotonic.
Note that large phase differences accumulate at very low velocities $v \sim 0.2 $
for even small black hole spins.}

\end{figure*}

\begin{figure*}
\includegraphics[]{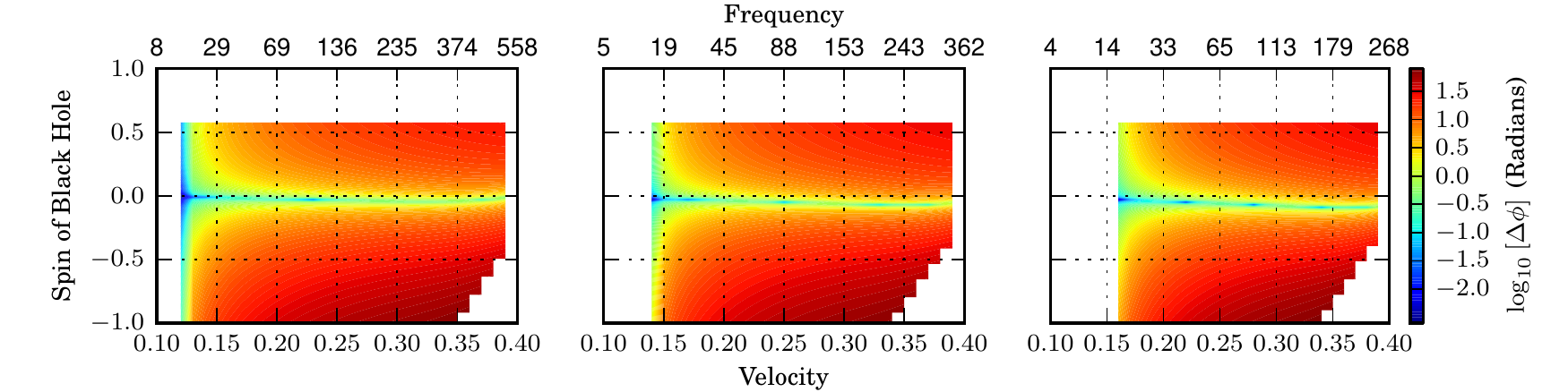}

\caption{\label{fig:T2SEpa} The accumulation of phase difference between
TaylorT2 and SEOBNRv1, for systems with component masses $(m_1, m_2)$ of
$(6\Msun, 1.4\Msun)$ (left), $(10\Msun, 1.4\Msun)$ (center), and $(14\Msun,
1.4\Msun)$ (right). TaylorT2 includes spin terms up to 2.5\ac{PN}.
The calculation starts from the velocity corresponding to a gravitational-wave  frequency of $15$Hz, 
continues to the velocity on the horizontal axis,
and reports the difference in accumulated gravitational-wave phase between the waveforms. The feature
in the bottom right corner of each plot arises because the TaylorT2 approximant is no longer monotonic.
As in Fig.~\ref{fig:T2T4pa}, a large phase difference is
accumulated at low velocities and small black hole spins.}

\end{figure*}

In the previous sections, we demonstrated that the two \ac{PN} approximants,
TaylorF2 and TaylorT4, and the SEOBNRv1 model are not faithful to each other.
We also showed that this is not due to the differences between frequency and
time domain waveforms.  From the construction of the TaylorR2F4 approximant, we
also demonstrated that the two \ac{PN} families can be written in a way that is
consistent up to the chosen \ac{PN} order, but where TaylorR2F4 contains higher
order in $v$ corrections that account for the differences between the models.
Since these are higher order corrections, they should start to become important
to the orbital phasing only at high velocities, and thus high  gravitational-wave 
frequencies. In this section we investigate where, for systems with parameters
corresponding to \ac{NSBH} binaries, the approximants diverge. We do this by
examining the accumulation of phase as a function of orbital velocity and
reporting the difference in the number of  gravitational-wave  cycles between different
approximants.

In Fig.~\ref{fig:T2T4pa}, we examine the difference in the accumulated phase
between TaylorT2 and TaylorT4 for three example systems with component masses
$(m_1, m_2)$ of $(6\Msun, 1.4\Msun)$, $(10\Msun, 1.4\Msun)$, and $(14\Msun,
1.4\Msun)$. We see that the phase difference between the two models quickly
grows to tens of radians, even when the black hole spin magnitude is small.
This is also true when comparing TaylorT2 and SEOBNRv1, as can be seen in
Fig.~\ref{fig:T2SEpa}.  In the latter case, there is also a noticeable
deviation away from the line of zero spin where for unknown reasons the two
models diverge and subsequently converge.

\section{Accumulation of mismatch}
\label{sec:faithfulness_match_accumulation}

\begin{figure*}
\includegraphics[]{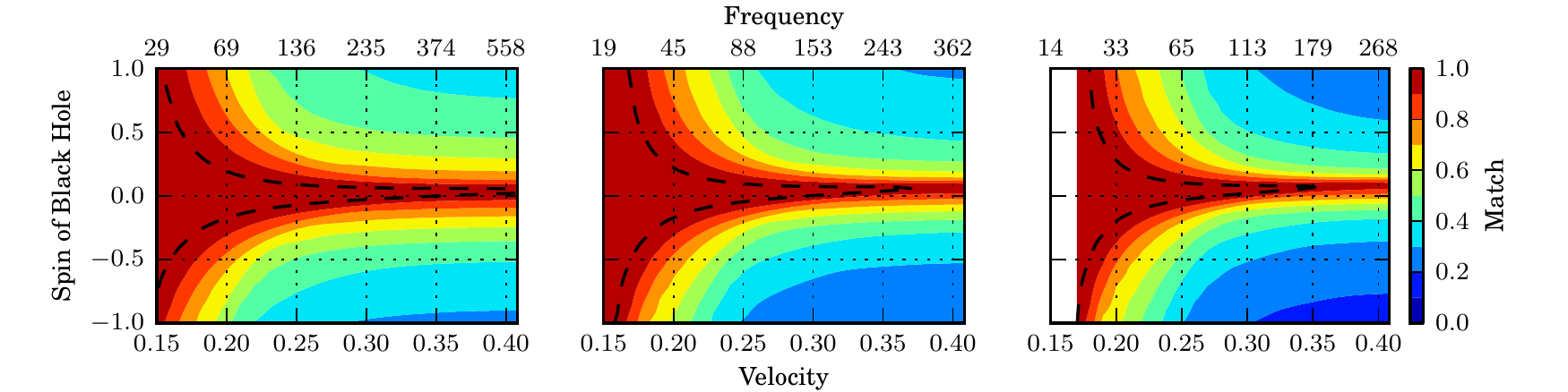}

\caption{\label{fig:F2T4ma} The match between TaylorF2 and TaylorT4
integrated from 15 Hz up to the designated frequency for systems with component
masses $(m_1, m_2)$ of $(1.4\Msun, 6\Msun)$ (left), $(1.4\Msun, 10\Msun)$
(center), and $(1.4\Msun, 14\Msun)$ (right).  Both approximants include spin corrections up to 2.5\ac{PN}.
Matches are calculated using the the aLIGO
zero-detuned, high-power sensitivity curve. A contour at a match of 0.97 is
indicated by the dotted line.  The match follows the general features seen in
the phase difference comparison of Fig.~\ref{fig:T2T4pa} and drops
significantly, even at relatively low velocities. For the $(1.4\Msun, 6\Msun)$ system with a black
hole spin $\chi = 0.5 $, the match has already dropped to $\sim 0.5$ at a velocity of only $\sim 0.25$.
}
\end{figure*}

\begin{figure*}
\includegraphics[]{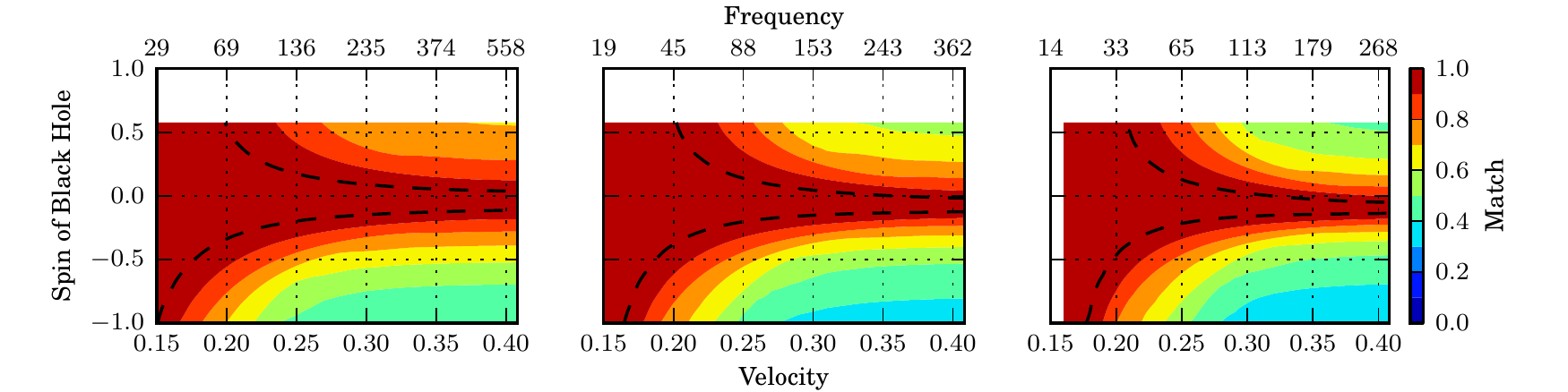}

\caption{\label{fig:F2SEma} The match between the TaylorF2 and SEOBNRv1 models
integrated from 15 Hz up to the designated frequency for systems with component
masses $(m_1, m_2)$ of $(6\Msun, 1.4\Msun)$ (left), $(10\Msun, 1.4\Msun)$
(center), and $(14\Msun, 1.4\Msun)$ (right). TaylorF2 includes spin corrections up to 2.5\ac{PN}.
Matches are calculated using the the aLIGO
zero-detuned, high-power sensitivity curve. A contour at a match of 0.97 is
indicated by the dotted line. We note that, although the match is marginally
improved compared to Fig.~\ref{fig:F2T4ma}, there are still large
disagreements at velocities as low as 0.25.}

\end{figure*}

As  gravitational-wave  detectors are not directly sensitive to phase differences alone, it
is useful to compute how the match, which incorporates the sensitivity of a
 gravitational-wave  detector, changes as a function of the upper frequency cutoff used for
the calculation. In this section we demonstrate at which frequencies and
corresponding velocities the match between waveform families drops. To do so,
we define an inner product between waveforms that is a function of the upper
frequency cutoff. This inner product is then used in the match calculation of
Eq.~\eqref{eq:match}.

In Fig.~\ref{fig:F2T4ma}, we examine the match between TaylorF2 and TaylorT4,
integrated from a lower frequency cutoff of 15 Hz up to the upper frequency
cutoff indicated on the horizontal axis. This is compared for the same three
example systems as in Sec.~\ref{sec:faithfulness_phase}. The match is shown
across the range of allowable values of the black hole spin and the neutron
star spin is set to zero. We see that the match drops precipitously even at low
velocities and relatively modest spin magnitudes. For example, for a system
with $m_1=6\Msun$, $m_2=1.4\Msun$, and a dimensionless spin of 0.5 for the
black hole, the match drops below 0.7 at a velocity of only 0.23. The loss in
match is more pronounced with increasing mass ratio. 

In Fig.~\ref{fig:F2SEma}, we examine the match between TaylorF2 and SEOBNRv1,
integrated from a lower frequency cutoff of 15Hz up to the upper frequency
cutoff indicated on the horizontal axis. Again, the match drops for large spin
magnitudes at relatively low velocities, although, just as the TaylorF2
approximant has shown better matches with the SEOBNRv1 approximant than with
the TaylorT4 approximant, this occurs at somewhat higher velocities. This shows
clearly that significant portions of the loss in match seen in
Sec.~\ref{sec:faithfulness} occurs at unexpectedly low velocities.

\section{Detection searches and Early aLIGO}
\label{sec:effectualness_and_flow}

\begin{figure}
\includegraphics{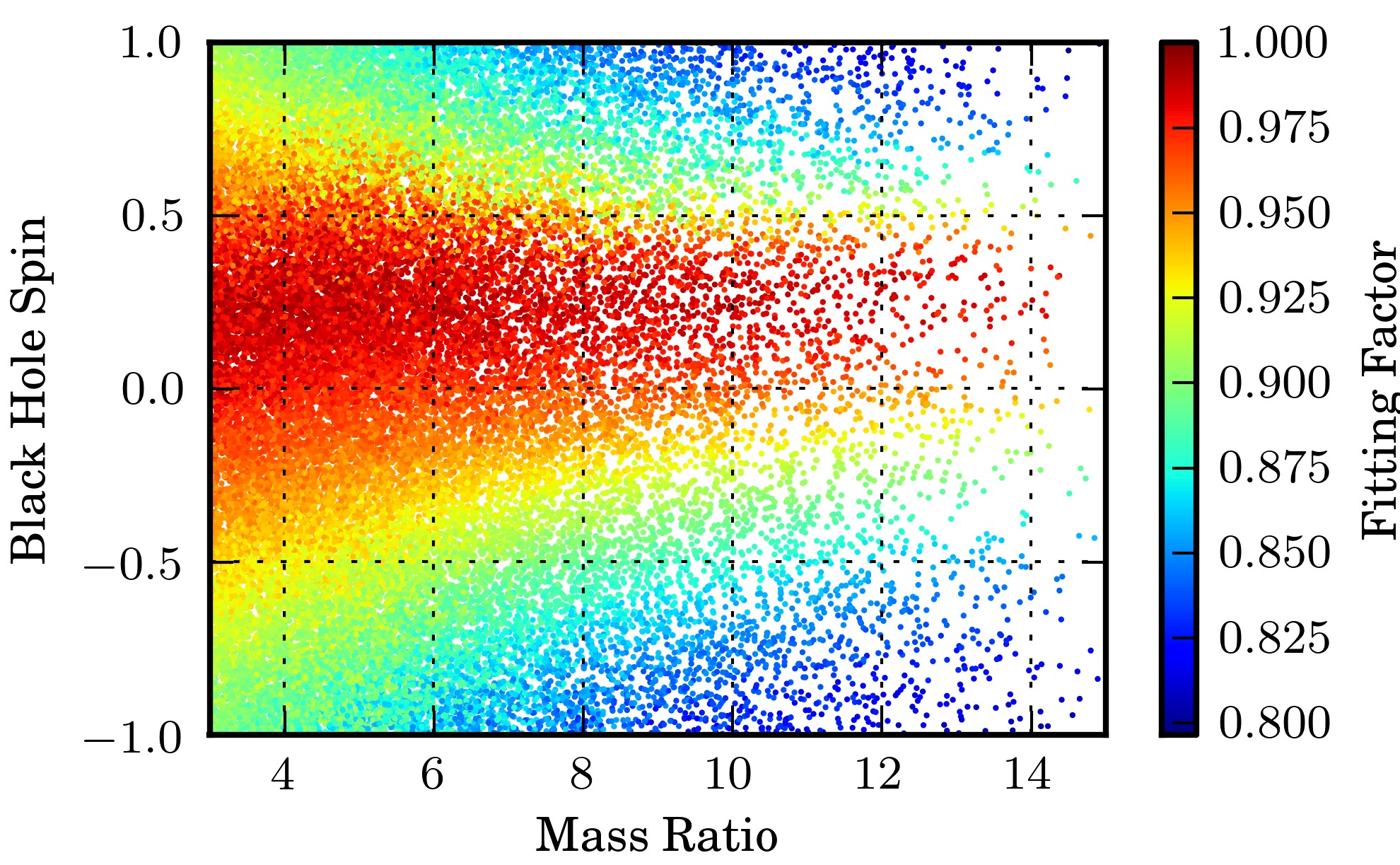}
\caption{\label{fig:spin2q}The fitting factor between the TaylorF2 and
TaylorT4 approximants as a function of the spin of the black hole
and the mass ratio of the system, when maximizing the match over a bank of
TaylorF2 waveforms. All approximants include spin corrections up to 2.5\ac{PN}.
Matches are calculated using the the aLIGO
zero-detuned, high-power sensitivity curve and a 15Hz lower frequency cutoff. In 
comparison to the match of these approximants shown in Fig.~\ref{fig:f2f4f}, we see that
while allowing for the maximization over a bank of templates has improved the overall agreement, 
it is unable to entirely make up for the poor match. 
}
\end{figure}

\begin{figure}
\includegraphics{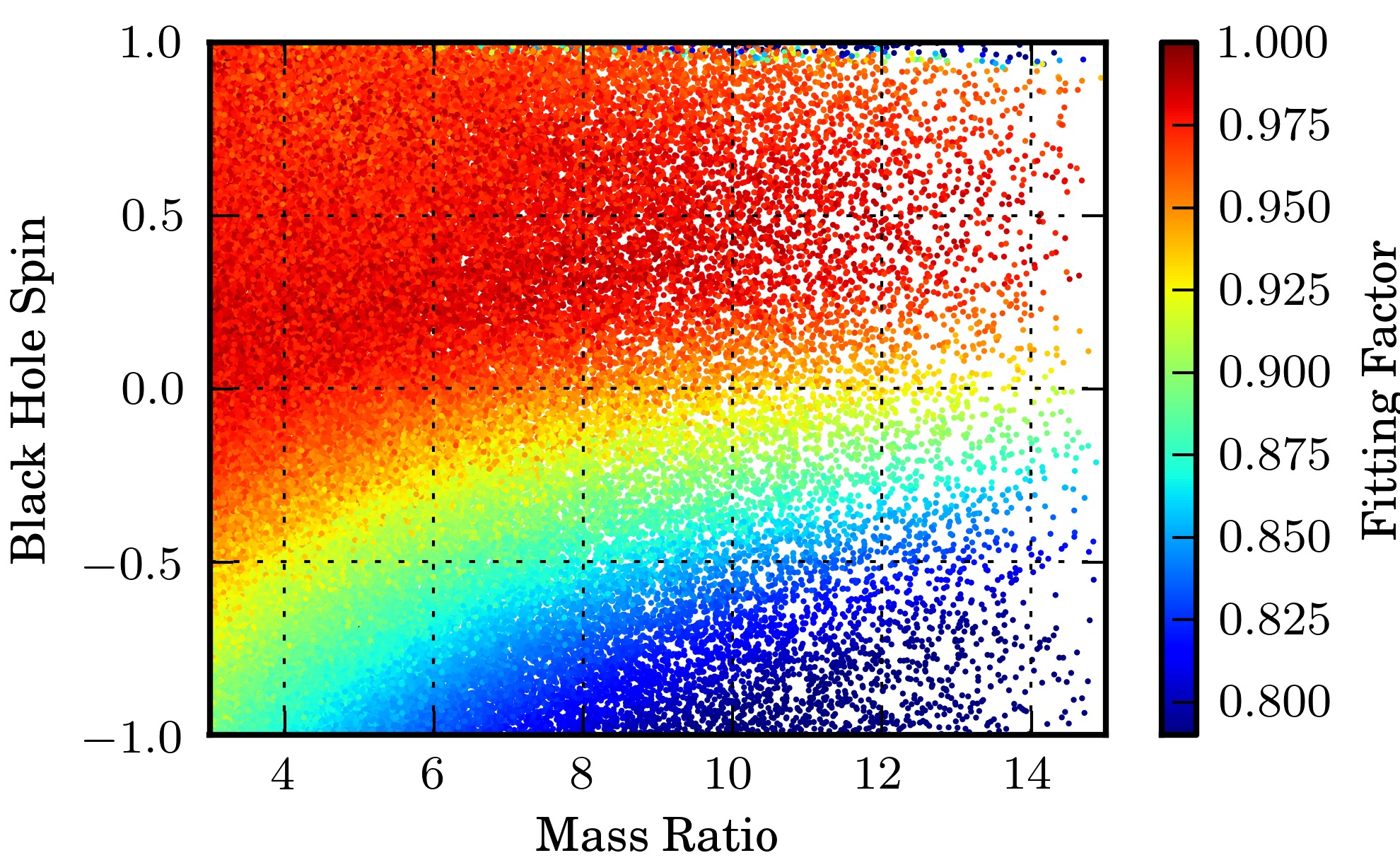}
\caption{\label{fig:spin2q7}The fitting factor between the TaylorF2 and
TaylorT4 approximants as a function of the spin of the black hole
and the mass ratio of the system, when maximizing the match over a bank of
TaylorF2 waveforms. All approximants include the 3.5\ac{PN} spin-orbit and 3.0\ac{PN} 
spin-orbit tail corrections. 
Matches are calculated using the the aLIGO
zero-detuned, high-power sensitivity curve and a 15Hz lower frequency cutoff. In 
comparison to the fitting factors shown in Fig.~\ref{fig:spin2q}, we see that adding
the higher order spin corrections has resulted in substantially improved fitting factors for 
systems where the spin is aligned with the orbital angular momentum. There is no 
improvement for anti-aligned systems.
}
\end{figure}

In the previous sections, we have demonstrated a substantial loss in match between
different \ac{PN} and EOB models of \ac{NSBH} binaries. These discrepancies
will cause substantial biases in attempts to measure the parameters of
detected systems with aLIGO. However, when detecting systems the
\emph{fitting factor}, rather than the match, is the quantity that is used to
assess the effectualness of a search~\cite{Apostolatos:1996rf}. The fitting
factor maximizes the match between a signal and a bank of templates designed
to capture e.g. $97\%$ of the optimal signal-to-noise ratio. The template bank is constructed to be 
valid for the same range of masses and spins used
throughout this paper and detailed in Sec~\ref{sec:introduction}. Furthermore, the mass and spin parameters of these templates
are sctrictly within this range. Discrepancies in
match due to differing approximants may be compensated for by allowing a waveform
to match to a template with shifted parameters.
Figs.~\ref{fig:spin2q} and~\ref{fig:spin2q7} show the fitting
factor of a TaylorF2 aligned spin template bank when used to detect aligned spin TaylorT4
waveforms. Fig.~\ref{fig:spin2q} shows the distribution of fittings factors for approximants that include up to the 2.5\ac{PN} 
spin corrections. Fig.~\ref{fig:spin2q7} demonstrates the effect of adding the higher order
3.0\ac{PN} spin-orbit tail and 3.5\ac{PN} spin-orbit corrections.
Construction of these aligned spin banks use the method introduced
in Ref.~\cite{Brown:2012qf} and is described in more detail in Ref.~\cite{Harry:2013effectualness}.

There is substantial improvement in the fitting factors of aligned spin systems when
adding the higher order spin corrections, but no improvement for anti-aligned spin systems. 
Although the loss in fitting factor is not as significant as the loss in match shown in
Figs.~\ref{fig:f2f4f} and~\ref{fig:f2t4fso}, aLIGO \ac{NSBH} searches will incur a significant loss in 
signal-to-noise ratio for systems with anti-aligned spins. Expanding the template bank may 
improve the fitting factors for particular systems. However, this would necessitate an increase
in the size of the template bank, and subsequently cause an increase
in the false alarm rate. If the faithfulness of \ac{NSBH} waveforms is improved, 
this approach is no longer necessary.

Refs.~\cite{Buonanno:2009zt} and ~\cite{Brown:2012nn} suggest a 12-solar-mass cutoff for
non-spinning, inpiral-only templates. While our figures consider a range of component masses, for a canonical
neutron star of 1.4 solar masses, this roughly corresponds to a mass ratio of 8. 
Although the fitting factors in Figs.~\ref{fig:spin2q} and~\ref{fig:spin2q7} are maximized over 
a parameter space that includes spin, we find that where the spin of the black hole is nearly zero,
the results are consistent with the comparisons of non-spinning approximants in Ref.~\cite{Buonanno:2009zt}.

\begin{figure}
\includegraphics{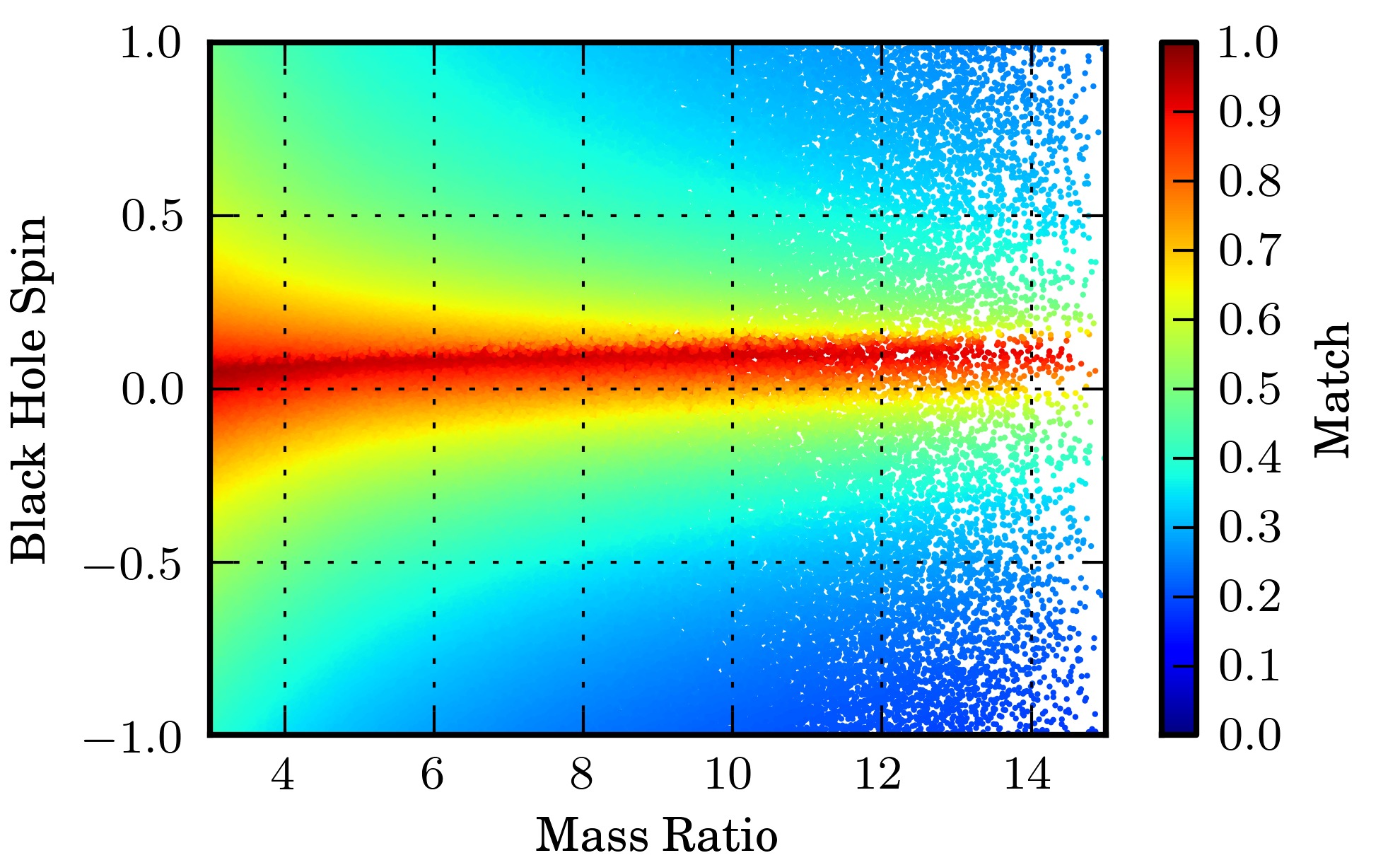}
\caption{\label{fig:5f2t430}The match between TaylorF2 and
TaylorT4 as a function of the spin of the black hole
and the mass ratio of the system. The approximants include spin corrections up to 2.5\ac{PN}. 
Matches are calculated using a 30Hz lower frequency cutoff to
approximate the sensitivity of an early \ac{aLIGO} detector. In comparison to Fig.~\ref{fig:f2f4f}, which uses a 15Hz lower
frequency cutoff, there is only a negligible improvement in match. Matches remain low at moderate black hole spins
$\chi \sim 0.3$.}
\end{figure}

\begin{figure}
\includegraphics{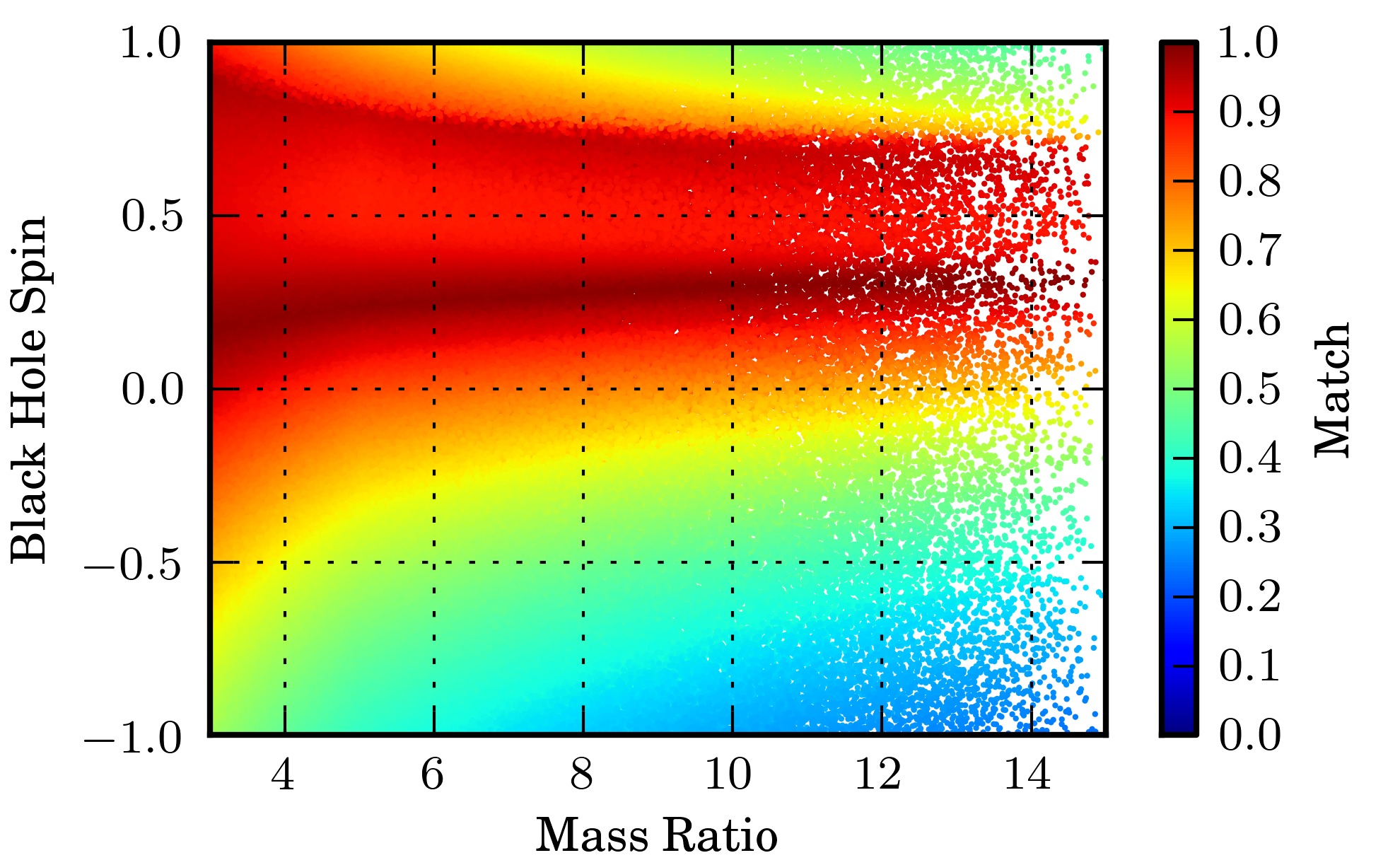}
\caption{\label{fig:7f2t430}The match TaylorF2 and
TaylorT4 approximants, with the 3.5\ac{PN} spin-orbit and 3.0\ac{PN} spin-orbit tail corrections
included, as a function of the spin of the black hole
and the mass ratio of the system.  The approximants include only the nown
spin terms up to 2.5\ac{PN}. Matches are calculated using a 30Hz lower frequency cutoff to
approximate the sensitivity of the early \ac{aLIGO} detector. In comparison to Fig.~\ref{fig:f2t4fso}, which uses a 15Hz lower
frequency cutoff, there is only a negligible improvement in match. }
\end{figure}

In the previous sections we have modeled the sensitivity of aLIGO
with the 
zero-detuned, high-power sensitivity curve~\cite{aLIGOSensCurves}. 
Early commissioning scenarios for \ac{aLIGO}
indicate that observations will begin with less sensitivity in the 10--40~Hz
region~\cite{Aasi:2013wya}. We investigate if the substantial disagreement found between TaylorF2 and TaylorT4 is still present for 
early detector sensitives by a instead using a lower frequency cutoff of 30 Hz. 

In Fig.~\ref{fig:5f2t430} and~\ref{fig:7f2t430}, we show the faithfulness between the TaylorF2 and TaylorT4 
approximants that include only the complete 2.5~\ac{PN} and partial 3.5\ac{PN} spin-related corrections, respectively. 
We see that there is no significant improvement in the faithfulness of the approximants,
and so additional spin corrections are desirable even for early detector scenarios. 

\vspace{0.5cm}
\section{Conclusions}
\label{sec:conclusion}

We have found that there is significant disagreement between \ac{NSBH}
waveforms modelled with the TaylorT2, TaylorT4, and SEOBNRv1 approximants. 
This will pose problems for the construction of optimal NSBH detection searches, 
potentially reducing the event rate, 
and may cause significant biases in the parameter measurement of detected signals.

The discrepancies are not accounted for by the differences between
frequency and time domain waveforms and start at fairly low ($v \sim 0.2$) orbital velocities.
Since the discrepancies in the approximants result from how the \ac{PN} expansions of the energy and flux
are combined and truncated, we conclude
that the calculation of higher order \ac{PN} terms is required to increase the
faithfulness of these approximants, and more importantly, to improve the
ability to detect \ac{NSBH} coalescences. The
discrepancies between approximants are significantly smaller when the spin of
the black hole is close to zero, which further motivates the calculation of the
\ac{PN} terms associated with the spin of the objects beyond those known
completely up to 2.5\ac{PN} order and partially up to 3.5\ac{PN}.
Therefore, additional work is needed to verify
the validity of waveform models used for \ac{NSBH} searches.
We also note that we have
only compared different waveform families under the assumption that the spins
of the component objects are (anti-)aligned with the orbital angular momentum
of the system.  It is expected that generic \ac{NSBH} systems will not be limited to
aligned spins, but may instead be more isotropically oriented.
This could lead to an additional source of discrepancy between our models and
the true signal, which would result in an additional loss in the detection rate
of sources.

\section*{Acknowledgements}
We thank Stefan Ballmer, Alessandra Buonanno, Eliu Huerta, Prayush Kumar, Richard O'Shaughnessy,
B.~S.~Sathyaprakash, Peter Saulson, and Matt West for useful discussions.  This work is supported
by National Science Foundation awards PHY-0847611 (DAB, AHN), PHY-1205835
(AHN, IWH), PHY-0970074 (EO), and PHY-0855589 (AL). 
DAB, IWH, AL, and EO thank the Kavli Institute for Theoretical Physics at
Santa Barbara University, supported in part by NSF grant PHY-0551164, for
hospitality during this work. DAB thanks the LIGO Laboratory Visitors Program, supported by NSF
cooperative agreement PHY-0757058, for hospitality.
DK and AL thank the Max
Planck Gesellschaft for support. DAB is supported by a Cottrell Scholar award
from the Research Corporation for Science Advancement.  Computations used in
this work were performed on the Syracuse University Gravitation and Relativity
cluster, which is supported by NSF awards PHY-1040231 and PHY-1104371.
\appendix
\onecolumngrid

\section{Post-Newtonian Energy and Gravitational-wave Flux}
\label{app:EF}

In this appendix, we give the \ac{PN} coefficients for the center of mass energy $E_i$ and
the  gravitational-wave  flux $F_i$, whose contributions were derived and presented
in~\cite{Damour:1999cr, Blanchet:2000nv, Blanchet:2001aw, Kidder:1995zr,
Faye:2006gx, Blanchet:2006gy, Poisson:1997ha, Mikoczi:2005dn, Arun:2008kb, 
Marsat:2012fn, Bohe:2012mr, Bohe:2013cla}.  We
include corrections that involve the component objects' spins up to 3.5\ac{PN}.
These coefficients depend on the dimensionless spins of the component objects
$\chi_i={\bf S}_i / m_i^2$, their projections onto the direction of so-called
Newtonian orbital angular momentum ${\bf L}_N= M \eta {\bf r} \times
\dot{\bf{r}}$, and the symmetric mass ratio $\eta$.  Additionally,
quadrupole-monopole contributions depend on a parameter $q_i$, which quantifies
the strength of the quadrupole moment induced by the oblateness of each
spinning compact object. For BHs, $q_i=1$, while for NSs $q_i$ will depend on
the equation of state, with~\cite{Laarakkers:1997hb} finding $q_i \sim 2 - 12$.

The coefficients associated with the energy are
given as
\begin{equation}
E_{\textrm{Newt}} = -\frac{M}{2}\eta,
\end{equation} \begin{equation}
E_2 = -\frac{3}{4} - \frac{1}{12}\eta,
\end{equation} \begin{equation}
E_3 = \sum_{i=1}^2 \left[ \frac{8}{3} \left(\frac{m_i}{M}\right)^2 + 2 \eta \right] \left( \chi_i \cdot \hat{L}_N \right) ,
\end{equation} \begin{equation}
E_4 = -\frac{27}{8} + \frac{19}{8}\eta - \frac{1}{24}\eta^2 
+ \eta \left[ \chi_1 \cdot \chi_2 - 3 \left( \chi_1 \cdot \hat{L}_N \right) \left(  \chi_2 \cdot \hat{L}_N \right) \right] 
+ \frac{1}{2} \sum_{i=1}^2 q_i \left( \frac{m_i}{M}\right)^2 \left[ \chi_i^2 - 3 \left( \chi_i \cdot \hat{L}_N \right)^2 \right] ,
\end{equation} \begin{equation}
E_5 = \sum_{i=1}^2 \left[ \left(8 - \frac{31}{9}\eta\right) \left(\frac{m_i}{M}\right)^2 
+ \eta \left( 3 - \frac{10}{3} \eta \right)\right] \left( \chi_i \cdot \hat{L}_N \right) ,
\end{equation} \begin{equation}
E_6 = -\frac{675}{64} + \left(\frac{34\,445}{576} -\frac{205}{96}
\pi^2\right)\eta - \frac{155}{96}\eta^2 -\frac{35}{5184}\eta^3 ,
\end{equation} \begin{equation}
E_7 = \sum_{i=1}^2 \left[ \left( 27 - \frac{211}{4}\eta + \frac{7}{6} \eta^2 \right) \left(\frac{m_i}{M}\right)^2
 + \eta \left( \frac{27}{4} - 39 \eta + \frac{5}{4}\eta^2\right) \right] \left( \chi_i \cdot \hat{L}_N \right) .
 \end{equation}

The coefficients associated with the flux are given as
\begin{equation}
F_{\textrm{Newt}} = \frac{32}{5}\eta^2,
\end{equation} \begin{equation}
F_2 = -\frac{1247}{336} - \frac{35}{12}\eta,
\end{equation} \begin{equation}
F_3 = 4 \pi - \sum_{i=1}^2 \left[ \frac{11}{4} \left(\frac{m_i}{M}\right)^2 
+ \frac{5}{4} \eta \right] \left( \chi_i \cdot \hat{L}_N \right) ,
\end{equation} \begin{eqnarray}
F_4 &=& -\frac{44\,711}{9072} + \frac{9271}{504}\eta + \frac{65}{18}\eta^2 
+ \eta \left[ \frac{289}{48} \left( \chi_1 \cdot \hat{L}_N\right) \left( \chi_2 \cdot \hat{L}_N\right) 
- \frac{103}{48} \chi_1 \cdot \chi_2 \right] \nonumber\\
 &+& \sum_{i=1}^2 q_i \left( \frac{m_i}{M}\right)^2 \left[ 3 \left( \chi_i \cdot \hat{L}_N \right)^2 - \chi_i^2 \right] 
 + \frac{1}{96}  \left( \frac{m_i}{M}\right)^2  \left[ 7 \chi_i^2 - \left( \chi_i \cdot \hat{L}_N \right)^2 \right] ,
\end{eqnarray} \begin{equation}
F_5 = \left( -\frac{8191}{672} - \frac{583}{24}\eta \right) \pi +  \sum_{i=1}^2 \left[ \left( -\frac{59}{16} 
+ \frac{701}{36}\eta\right) \left(\frac{m_i}{M}\right)^2 + \eta \left( -\frac{13}{16} 
+ \frac{43}{4}\eta\right) \right] \left( \chi_i \cdot \hat{L}_N \right) ,
\end{equation} \begin{eqnarray}
F_6 &=& \frac{6\,653\,739\,519}{69\,854\,400} + \frac{16}{3}\pi^2 - \frac{1712}{105}
\gamma_E -\frac{856}{105} \log(16v^2) + \left(-\frac{134\,543}{7776} +
\frac{41}{48}\pi^2\right)\eta \nonumber\\
 &-& \frac{94\,403}{3024}\eta^2 - \frac{775}{324}\eta^3
 - \pi \sum_{i=1}^2 \left[ \frac{65}{6} \left(\frac{m_i}{M}\right)^2 - \frac{31}{6}\eta \right] \left( \chi_i \cdot \hat{L}_N \right),
\end{eqnarray} \begin{eqnarray}
F_7 &=& \left( -\frac{16\,285}{504} + \frac{214\,745}{1728}\eta +\frac{193\,385}{3024}\eta^2\right)\pi \nonumber\\
 &+& \sum_{i=1}^2 \left[ \left( \frac{162\,035}{3888} + \frac{971}{54}\eta - \frac{6737}{108}\eta^2 \right) \left(\frac{m_i}{M}\right)^2 + \eta \left( \frac{9535}{336} + \frac{1849}{126}\eta - \frac{1501}{36}\eta^2 \right) \right] \left( \chi_i \cdot \hat{L}_N \right) .
\end{eqnarray}

\section{Post-Newtonian Approximants}

The PN approximants TaylorT4, TaylorT2, TaylorF2, and TaylorR2F4 are given
using the flux up to 3.5 PN and the center-of-mass energy up to 3.0 PN.
Corrections due to spin are included up to 3.5 PN order. This includes the
leading order spin orbit correction $\beta$ at 1.5PN, leading order spin-spin correction $\sigma$ at 2PN 
(which includes quadrupole-monopole and so-called self-spin effects proportional to $s_i^2$), 
next-to-leading order spin-orbit corrections $\gamma$ at 2.5PN, tail-induced spin orbit correction $\xi$ at 3PN, 
and third order spin-orbit correction $\zeta$ appearing at 3.5PN. These corrections can be expressed as,
\begin{eqnarray}
\beta =  \sum_{i=1}^2 \left[ \frac{113}{12} \left(\frac{m_i}{M}\right)^2 + \frac{25}{4}\eta \right] \left(\chi_i \cdot
\hat{L}_N\right) ,
\end{eqnarray}

\begin{eqnarray}
\sigma = \eta \left[ \frac{721}{48} \left( \chi_1 \cdot \hat{L}_N\right) \left( \chi_2 \cdot \hat{L}_N\right) 
- \frac{247}{48} \chi_1 \cdot \chi_2 \right]\nonumber\\
+\frac{5}{2} \sum_{i=1}^2 q_i \left( \frac{m_i}{M}\right)^2 
\left[ 3 \left( \chi_i \cdot \hat{L}_N \right)^2 - \chi_i^2 \right] \nonumber\\
+ \frac{1}{96} \sum_{i=1}^2 \left( \frac{m_i}{M}\right)^2 \left[ 7 \chi_i^2 - \left( \chi_i \cdot \hat{L}_N \right)^2 \right] ,
\end{eqnarray}

\begin{eqnarray}
 \gamma = \sum_{i=1}^2 \left[  \left( \frac{732\,985}{2268} + \frac{140}{9}\eta\right) \left(\frac{m_i}{M}\right)^2 
 + \eta \left(\frac{13\,915}{84} - \frac{10}{3}\eta \right) \right] \left( \chi_i \cdot \hat{L}_N\right) ,
 \end{eqnarray}

\begin{eqnarray}
\xi = \pi \sum_{i=1}^2 \left[ \frac{75}{2} \left(\frac{m_i}{M}\right)^2 + \frac{151}{6} \eta \right] 
\left( \chi_i \cdot \hat{L}_N \right) ,
\end{eqnarray}

\begin{eqnarray}
\zeta = \sum_{i=1}^2 \left[ \left( \frac{130\,325}{756} - \frac{796\,069}{2016}\eta + \frac{100\,019}{864}\eta^2\right) 
\left(\frac{m_i}{M}\right)^2 + \eta \left(\frac{1\,195\,759}{18\,144} - \frac{257\,023}{1008}\eta + \frac{2903}{32}\eta^2\right) 
\right] \left( \chi_i \cdot \hat{L}_N \right) .
\end{eqnarray}

\subsection{TaylorT4}
\label{app:T4}
\begin{eqnarray} \label{eq:dvdtT4}
\frac{dv}{dt} = \frac{32 \eta}{5 M} v^9 \Bigg\{
1 &+& 
\left( -\frac{743}{336} - \frac{11}{4} \eta \right) v^2
+ (4 \pi - \beta) v^3
+ \left( \frac{34\,103}{18\,144} + \frac{13\,661}{2016} \eta + \frac{59}{18} \eta^2 + \sigma \right) v^4
\nonumber \\
&+& \left( -\frac{4159 \pi}{672} -\frac{189 \pi}{8} \eta - \frac{9}{40} \gamma + \Big(\frac{743}{168} + \frac{11}{2} \eta \Big) \beta \right) v^5
\nonumber \\
&+& \left[ \frac{16\,447\,322\,263}{139\,708\,800} - \frac{1712 \gamma_E}{105} + \frac{16 \pi^2}{3} - \frac{1712}{105} \log (4 v) \right.
\nonumber \\
& & \qquad + \left. \Big(-\frac{56\,198\,689}{217\,728} + \frac{451 \pi ^2}{48} \Big) \eta + \frac{541}{896} \eta^2 - \frac{5605}{2592} \eta^3 - \xi \right] v^6 \nonumber \\
&+& \pi \left( -\frac{4415}{4032} + \frac{358\,675}{6048} \eta + \frac{91\,495}{1512} \eta^2 - \zeta \right) v^7
\Bigg\}
\end{eqnarray}

\subsection{TaylorT2}
\label{app:T2}
\begin{eqnarray}
\frac{dt}{dv} = \frac{5 M}{32 \eta} v^{-9} \Bigg\{
1 &+& 
\left(\frac{743}{336} + \frac{11}{4} \eta \right) v^2
+ \left( -4 \pi + \beta \right) v^3
+ \left( \frac{3\,058\,673}{1\,016\,064} + \frac{5429}{1008} \eta + \frac{617}{144} \eta^2 - \sigma \right) v^4
\nonumber \\
&+& \left( -\frac{7729 \pi}{672} + \frac{13 \pi}{8} \eta + \frac{9}{40} \gamma \right) v^5
\nonumber \\
&+& \Bigg[ -\frac{10\,817\,850\,546\,611}{93\,884\,313\,600} + \frac{32 \pi^2}{3} + \frac{1712 \gamma_E}{105} + \frac{1712}{105} \log (4 v)
\nonumber \\
& & \qquad + \Big(\frac{3\,147\,553\,127}{12\,192\,768} - \frac{451 \pi^2}{48} \Big) \eta - \frac{15\,211}{6912} \eta^2 
+ \frac{25\,565}{5184} \eta^3 - 8 \pi \beta + \xi \Bigg] v^6 \nonumber \\
&+& \pi \Bigg( -\frac{15\,419\,335}{1\,016\,064} -\frac{75\,703}{6048} \eta + \frac{14\,809}{3024} \eta^2 
- \beta \left( \frac{8\,787\,977}{1\,016\,064} + \frac{51\,841}{2016}\eta + \frac{2033}{144}\eta^2 \right) \nonumber \\
 + \gamma \left( \frac{2229}{2240} + \frac{99}{80}\eta \right) + \zeta \Bigg) v^7
\Bigg\}
\end{eqnarray}

\subsection{TaylorF2}
\label{app:F2}

\begin{equation}
A_{(F2)}(f) \propto \frac{\left(\pi M f\right)^{2/3}}{\sqrt{\dot{F}(f)}}  
\end{equation} \begin{eqnarray}
\psi_{(F2)}(f) &=& 2\pi f t_c - \phi_c + \frac{3}{128\eta} v^{-5} \Bigg\{
1
+ \left( \frac{3715}{756} + \frac{55}{9} \eta \right) v^2 
+ ( 4 \beta - 16\pi ) v^3 \nonumber\\
&+& \left( \frac{15\,293\,365}{508\,032} + \frac{27\,145}{504} \eta + \frac{3085}{72} \eta^2 - 10\sigma \right) v^4
+ \left( \frac{38\,645}{756} \pi - \frac{65}{9}\pi\eta - \gamma \right) (1+3\log(v)) ~ v^5\nonumber\\
&+& \Bigg[ \frac{11\,583\,231\,236\,531}{4\,694\,215\,680} - \frac{640}{3}\pi^2 -\frac{6848\gamma_E}{21} -\frac{6848}{21} \log(4v)
+ \left( - \frac{15\,737\,765\,635}{3\,048\,192} + \frac{2255\pi^2}{12} \right) \eta \nonumber\\
&+& \frac{76\,055}{1728} \eta^2 - \frac{127\,825}{1296} \eta^3 + 160 \pi \beta - 20 \xi \Bigg] v^6 + \pi \left( \frac{77\,096\,675}{254\,016} + \frac{378\,515}{1512}\eta - \frac{74\,045}{756}\eta^2 
\right. \nonumber\\
 &+& \left. \beta \left( \frac{43\,939\,885}{254\,016} + \frac{259\,205}{504}\eta + \frac{10\,165}{36}\eta^2 \right) 
 - \gamma \left( \frac{2229}{112} - \frac{99}{4}\eta \right) - 20 \zeta \right) v^7  \Bigg\}
\end{eqnarray}

\subsection{TaylorR2F4}
\label{app:R2F4}

In the equation below, the $a_i$ are the PN coefficients of the TaylorT4 expansion 
\begin{equation}
\left(\frac{dv}{dt}\right)_{T4} = A_7(v) = a_0 \left( 1 + \sum_{i=2}^7 a_i v^i + a_{6\log} v^6 \log(4 v)\right) . \nonumber
\end{equation}
which can be read off of Eq.~(\ref{eq:dvdtT4}).

\begin{eqnarray}
\psi_{(R2F4)}(f) &=& \psi_{(F2)}(f) + \frac{3}{128\eta}v^{-5} \Bigg\{ \bigg[ - 20 \beta^2
 + \sigma \left(\frac{3715}{42} + 100 \eta \right)\bigg] v^6 
 + \bigg[ \left( 40 \beta - 160 \pi \right) \sigma \bigg]  v^7 \nonumber\\
&+& \frac{40}{9} \bigg[ \left( 3 a_2^2 a_4 - a_2^4 - a_4^2 - 2 a_3 a_5+ 3 a_2 a_3^2 - 2 a_2 a_6 
+ \frac{2}{3} a_2 a_{6\log} \right) \left(1 - 3 \log(v)\right)\nonumber\\ 
&+& 3 a_2 a_{6\log} \log(4v)^2\bigg] v^8 
+ 5 \bigg[ 8 a_2^2 a_3 - 2 a_3^3 - 12 a_2 a_3 a_4 - 6 a_2^2 a_5 + 4 a_4 a_5 + 4 a_3 a_6 - 5 a_3 a_{6\log}\nonumber\\ 
&+& 4 a_2 a_7 + 4 a_3 a_{6\log} \log(4v)\bigg] v^9
+ 4 \bigg[ -a_2^5 + 6 a_2^2 a_3^2 + 4 a_2^3 a_4 - 3 a_3^2 a_4 - 3 a_2 a_4^2 - 6 a_2 a_3 a_5\nonumber\\ 
&+& a_5^2 - 3 a_2^2 a_6 + 2 a_4 a_6 + 2 a_3 a_7 + \left(\frac{21}{10} a_2^2 - \frac{7}{5} a_4\right) a_{6\log} 
+ \left( 2 a_4 a_{6\log} - 3 a_2^2 a_{6\log} \right) \log(4v) \bigg] v^{10}\nonumber\\
&+& \frac{20}{9} \bigg[ -5 a_2^4 a_3 + 4 a_2 a_3^3 + 4 a_2^3 a_5 
+ 12 a_2^2 a_3 a_4 - 3 a_3 a_4^2 - 3 a_3^2 a_5 - 6 a_2 a_4 a_5 - 6 a_2 a_3 a_6 \nonumber\\
&+& 2 a_5 a_6 + 3 a_2 a_3 a_{6\log} - a_5 a_{6\log} + 2 a_4 a_7 - 3 a_2^2 a_7 
+ \left(2 a_5 - 6 a_2 a_3 \right) a_{6\log} \log(4v)\bigg] v^{11} \nonumber\\
&+& \frac{10}{7} \bigg[ a_2^6 - 10 a_2^3 a_3 ^2 
+ a_3^4 - 5 a_2^4 a_4 + 12 a_2 a_3^2 a_4 + 6 a_2^2 a_4^2
 - a_4^3 + 12 a_2^2 a_3 a_5 - 6 a_3 a_4 a_5 - 3 a_2 a_5^2 \nonumber\\
 &+&\bigg. 4 a_2^3 a_6 - 3 a_3^2 a_6 - 6 a_2 a_4 a_6 + a_6^2 + \left( -\frac{11}{7} a_2^3 + \frac{33}{28} a_3^2 + \frac{33}{14} a_2 a_4 
 - \frac{11}{14} a_6 + \frac{93}{392} a_{6\log} \right) a_{6\log} - 6 a_2 a_3 a_7\nonumber\\ 
 &+& 2 a_5 a_7 + \left( 4 a_2^3 - 3 a_3^2 - 6 a_2 a_4 + 2 a_6 
 - \frac{11}{14} a_{6\log} \right) a_{6\log} \log(4v) + a_{6\log}^2 \log(4v)^2 \bigg] v^{12} \Bigg\}
\end{eqnarray}

\bibliography{references}

\end{document}